\newif\ifconfver
\confverfalse      
\confvertrue        

\newif\ifplainver  
\plainvertrue

\newif\ifhide  
\hidetrue

\ifplainver
    \confverfalse   
\fi

\ifconfver
     \documentclass[10pt,twocolumn,twoside]{IEEEtran}
\else
    \ifplainver
        \documentclass[11pt]{article}
        \usepackage{fullpage}
    \else
        \documentclass[12pt,draftcls,onecolumn]{IEEEtran}
    \fi
\fi

\usepackage{etoolbox,soul}%
\usepackage{xpatch}
\usepackage{blindtext}
\usepackage{tocloft}%
\newlength{\articlesectionshift}%
\let\LaTeXStandardSection\section
\let\LaTeXStandardTheSection\thesection
\let\LaTeXStandardTheSubSection\thesubsection
\let\LaTeXStandardTheSubSubSection\thesubsubsection
\let\LaTeXStandardTheParagraph\theparagraph

\makeatletter
\newcounter{titlecounter}

\xpretocmd{\maketitle}{\ifnumgreater{\value{titlecounter}}{1}}{\clearpage}{}{} 
\xpatchcmd{\maketitle}{\let\maketitle\relax\let\@maketitle\relax}{\refstepcounter{titlecounter}\begingroup
  \addtocontents{toc}{\begingroup\addtolength{\cftsecindent}{-\articlesectionshift}}%
  \addcontentsline{toc}{section}{\protect{\numberline{\thetitlecounter}{\@title~ \@author}}}%
  \addtocontents{toc}{\endgroup}
}{%
  \typeout{Patching was successful}
}{%
  \typeout{patching failed}
}%

\def\@IEEEdestroythesectionargument#1{\LaTeXStandardSection{#1}}%

\xapptocmd{\maketitle}{%
\renewcommand{\thesection}{\LaTeXStandardTheSection}%
\renewcommand{\thesubsection}{\LaTeXStandardTheSubSection}%
\renewcommand{\thesubsubsection}{\LaTeXStandardTheSubSubSection}%
\renewcommand{\theparagraph}{\LaTeXStandardTheParagraph}%
}{}{}%

\@addtoreset{section}{titlecounter}

\usepackage{calc,amsfonts,amssymb,amsmath,bm,url,color,theorem,graphicx,cite}
\usepackage{psfrag,float}
\usepackage{algorithm}
\usepackage{algorithmic}
\usepackage{ marvosym }
\usepackage{enumerate}
\usepackage{bbm}
\usepackage{shortcuts_OPT}
\usepackage{multirow}
\usepackage{subcaption}
\usepackage{diagbox}

\makeatletter
\newcommand\figcaption{\def\@captype{figure}\caption}
\newcommand\tabcaption{\def\@captype{table}\caption}
\makeatother

\usepackage{eqparbox}

\definecolor{orange}{RGB}{255,107,0}

\newtheorem{Fact}{Fact}

\newtheorem{Prop}{Proposition}

\newtheorem{Corollary}{Corollary}

\newtheorem{Heuristic}{Heuristic}

\theorembodyfont{\rmfamily}

\newtheorem{Remark}{Remark}

\newcommand\bx{\ensuremath{{\bm x}}}
\newcommand\by{\ensuremath{{\bm y}}}

\newcommand\bz{\ensuremath{{\bm z}}}
\newcommand\bp{\ensuremath{{\bm p}}}
\newcommand\bP{\ensuremath{{\bm P}}}
\newcommand\bR{\ensuremath{{\bm R}}}

\newcommand\bX{\ensuremath{{\bm X}}}

\newcommand\bC{\ensuremath{{\bm C}}}
\newcommand\bc{\ensuremath{{\bm c}}}
\newcommand\ba{\ensuremath{{\bm a}}}

\newcommand\bA{\ensuremath{{\bm A}}}
\newcommand\bbA{\ensuremath{\bar{\bm A}}}
\newcommand\bb{\ensuremath{{\bm b}}}

\newcommand\bB{\ensuremath{{\bm B}}}

\newcommand\bmu{\ensuremath{{\bm \mu}}}

\newcommand\bd{\ensuremath{{\bm d}}}
\newcommand\bD{\ensuremath{{\bm D}}}
\newcommand\bu{\ensuremath{{\bm u}}}
\newcommand\bv{\ensuremath{{\bm v}}}
\newcommand\bSig{\ensuremath{{\bm \Sigma}}}

\newcommand\bY{\ensuremath{{\bm Y}}}

\newcommand\bU{\ensuremath{{\bm U}}}
\newcommand\bs{\ensuremath{{\bm s}}}
\newcommand\bS{\ensuremath{{\bm S}}}

\newcommand{\Rbb}{\mathbb{R}}

\newcommand{\setB}{\mathcal{B}}

\newcommand{\setE}{\mathcal{E}}

\newcommand{\setX}{\mathcal{X}}

\newcommand{\setU}{\mathcal{U}}

\newcommand{\setS}{\mathcal{S}}

\newcommand{\setN}{\mathcal{N}}

\newcommand{\Exp}{\mathbb{E}}

\newcommand{\Diag}{\mathrm{Diag}}

\newcommand{\bzero}{{\bm 0}}
\newcommand{\bone}{{\bm 1}}
\newcommand{\bI}{{\bm I}}

\newcommand\indfn[1]{{{\mathbbm 1}_{#1}}}

\newcommand\conv{\ensuremath{{\rm conv}}}

\newcommand\svol{\ensuremath{{\rm vol}}}

\usepackage{tikz}
\usetikzlibrary{arrows.meta}
\usetikzlibrary{decorations}
\usetikzlibrary{calc}
\usetikzlibrary{shapes.geometric}
\usetikzlibrary{external}

\begin{document}


\newcommand{\papertitle}{
SISAL Revisited
}

\newcommand{\paperabstract}{
Simplex identification via split augmented Lagrangian (SISAL) is a popularly-used algorithm in blind unmixing of hyperspectral images.
Developed by Jos\'e M. Bioucas-Dias in 2009, the algorithm is fundamentally relevant to tackling simplex-structured matrix factorization, and by extension, non-negative matrix factorization, which have many applications under their umbrellas.
In this article, we revisit SISAL and provide new meanings to this quintessential algorithm.
The formulation of SISAL was motivated from a geometric perspective, with no noise.
We show that SISAL can be explained as an approximation scheme from a probabilistic simplex component analysis framework, which is statistical and is principally more powerful in accommodating the presence of noise.
The algorithm for SISAL was designed based on a successive convex approximation method, with a focus on practical utility.
It was not known, by analyses, whether the SISAL algorithm has any kind of guarantee of convergence to a stationary point.
By establishing associations between the SISAL algorithm and a line-search-based proximal gradient method, we confirm that SISAL can indeed guarantee convergence to a stationary point.
Our re-explanation of SISAL also reveals new formulations and algorithms.
The performance of these new possibilities is demonstrated by numerical experiments.
}


\ifplainver


\title{\papertitle}

\author{
Chujun Huang$^\dag$*, Mingjie Shao$^\dag$*, Wing-Kin Ma$^\dag$, and Anthony Man-Cho So$^\S$ \\ ~ \\
	$^\dag$Department of Electronic Engineering, The Chinese University of Hong Kong, \\
	Hong Kong SAR of China \\ ~ \\
	$^\S$Department of Systems Engineering and Engineering Management, \\ The Chinese University of Hong Kong,
	Hong Kong SAR of China \\ ~ \\
}

\maketitle

\begin{abstract}
	\paperabstract
\end{abstract}

\else
\title{\papertitle}

\ifconfver \else {\linespread{1.1} \rm \fi
	
	\author{
		Wing-Kin Ma
	}

	\maketitle
	
	\ifconfver \else
	\begin{center} \vspace*{-2\baselineskip}
	\end{center}
	\fi
	
	\begin{abstract}
		\paperabstract
	\end{abstract}
	
	
	\begin{IEEEkeywords}\vspace{-0.0cm}
		Simplex component analysis, hyperspectral unmixing, maximum
		likelihood, simplex volume minimization
	\end{IEEEkeywords}
	
	\ifconfver \else \IEEEpeerreviewmaketitle} \fi

\fi

\ifconfver \else
\ifplainver \else
\newpage
\fi \fi


\section{Introduction}

{\let\thefootnote\relax\footnotetext{{This work was supported by a General Research Fund of Hong Kong Research Grant Council under Project ID CUHK 14205717.}}}
{\let\thefootnote\relax\footnotetext{{*Chujun Huang and Mingjie Shao contributed equally to this work.}}}

Simplex identification via split augmented Lagrangian (SISAL) is an algorithm developed by Jos\'e M. Bioucas-Dias in 2009 \cite{Dias2009}.
It appears in a $4$-page conference paper, with open source code (in MATLAB).
It basically deals with a simplex-structured matrix factorization problem from  hyperspectral imaging; the problem is famously known as hyperspectral unmixing (HU) in the community of hyperspectral remote sensing.
It is worth mentioning that HU is not only a key topic in hyperspectral imaging \cite{Jose12,Ma2014HU}, it also has strong relationships with non-negative matrix factorization and the various machine learning applications thereof; see, e.g., \cite{fu2019nonnegative,gillis2021nmf} and the references therein.
The development of SISAL revolves around problem formulation and optimization algorithm design.
SISAL has a unique place in the course of history of HU:
it offered one of the first, and most pioneering, practical algorithms for a promising but difficult-to-implement strategy for HU, namely, simplex volume minimization (SVMin).
It has become a benchmark and has been frequently used by researchers.
By the authors' understanding, the reasons boil down to one: {\em it works well in practice.}
SISAL has good running speed, scales well with the data sizes (very large ones) computationally, delivers reasonably good unmixing results, and demonstrates resilience to noise and modeling error effects.
SISAL shows powerful intuitions by its inventor. As an article to pay tribute to Bioucas-Dias' tremendous insights to hyperspectral imaging, allow us to quote a saying by Steve Jobs: ``Intuition is a very powerful thing, more powerful than intellect, in my opinion.''

This article serves as an endeavor to continue the legacy of Bioucas-Dias' SISAL.
It can also be regarded as the sequel of \cite{PRISM2021}.
The SISAL work has left some open questions.
First and foremost, SISAL requires tuning of a regularization parameter.
That parameter has an impact on SISAL's noise resilience behaviors.
It is not clear how we should choose that parameter,
apart from empirical or human experience.
To make the story more complicated, SISAL was motivated by the noiseless case, and the subsequent explanation of why SISAL works in the noisy case was intuitive.
Our question is whether there exists an alternative explanation for the noisy case.
To answer that, we pursue a probabilistic simplex component analysis (SCA) framework, wherein we employ a principled formulation, namely, the maximum likelihood, to deal with the problem under a pertinent statistical model (to be specified later).
This statistical strategy for unmixing is different from SISAL or SVMin, which is geometric.
The former, by principle, has the upper hand in the noisy case; it also frees us from  parameter tuning.
We will show that SISAL can be seen as an approximation scheme of probabilistic SCA.
Moreover, the connections we build suggest a different concept: Rather than considering parameter tuning, we should work on a more general formulation of SISAL, which is induced from probabilistic SCA and has no pre-selected parameter (except for the noise variance which can be estimated from data).

Some prior work on the aforementioned direction should be recognized.
The links between SVMin (but not SISAL) and statistical inference were noted in earlier works \cite{nascimento2009learning,nascimento2012hyperspectral}, \cite[Appendix]{dobigeon2009joint}.
The prequel of this article \cite{PRISM2021} describes the connections between SVMin and probabilistic SCA more explicitly, but it only showed similarities, not a direct connection, between SISAL and probabilistic SCA.
This article shows a close connection between SISAL and probabilistic SCA, compared to the previous work.
Curiously, a simple second-order statistics observation (to be shown in Section~\ref{sect:prism_con}) provides the very crucial piece of jigsaw to complete the puzzle.

Second, it is intriguing to study the optimization aspects of SISAL.
The problem formulated in SISAL is non-convex, and
Bioucas-Dias derived a successive convex approximation algorithm to tackle the problem.
The algorithm can be seen a first-order method, as will be elaborated upon later, and it is worth mentioning that, in 2009, non-convex first-order optimization was not as extensively studied as today.
As mentioned, the algorithm proved to be a success in practice.
Our question is whether the SISAL algorithm actually possesses any form of guarantees of finding a stationary point, leveraging on our much better understanding of non-convex first-order optimization today.
We will see that the SISAL algorithm can be viewed as an instance of the proximal gradient method, with line search along the feasible direction.
There are, however, caveats that prevent us from directly claiming convergence to a stationary point---a key component in the objective function does not have Lipschitz gradient, and its domain is the set of all invertible matrices (which is a non-convex set).
In this connection we should mention that, in the current non-convex first-order optimization literature, it is very common to assume the aforementioned component to have Lipschitz gradient.
We will confirm that the SISAL algorithm, with a minor adjustment, can indeed guarantee convergence to a stationary point (more accurately, limit-point convergence).
This is made possible by establishing associations between the SISAL algorithm and the line-search-based proximal gradient framework in \cite{BLPP16}.

Our endeavor to re-explain SISAL also gives rise to new insights for algorithms.
Through connecting SISAL and probabilistic SCA, we see a more general formulation that resembles SISAL.
The new formulation replaces SISAL's penalty term with a probabilistic penalty term, and it has the regularization parameter (which requires tuning in SISAL) eliminated.
We custom-design a practical algorithm for the formulation (which is more difficult than the SISAL), and we will illustrate by numerical experiments that this probabilistic SISAL performs well under the high SNR regime.
We also study a SISAL variant that is easier to work with from an optimization algorithm design viewpoint, and numerical results suggest that the variant is computationally competitive.

We organize this paper as follows.
Section 2 provides the problem statement and reviews the formulation of SISAL.
Section 3 studies probabilistic SCA, shows how probabilistic SCA and SISAL are connected, and, in the process, reveals new formulations.
Section 4 considers the optimization aspects of SISAL, particularly, the stationarity guarantee of SISAL.
Section 5 develops a practical algorithm for the new formulation of probabilistic SISAL.
Section 6 provides synthetic and semi-real data experiments.
Section 7 concludes this work.

Our basic notations are as follows.
The sets of all real, non-negative and positive numbers are denoted by $\Rbb, \Rbb_+, \Rbb_{++}$, respectively;
boldface lowercase letters, such as $\bx$, represent column vectors;
boldface capital letters, such as $\bX$, represent matrices;
we may use the notation $(x_1,\ldots,x_n)$ to represent a column vector;
the superscripts $^\top$, $^{-1}$ and $^\dag$ denote transpose, inverse and pseudo-inverse, respectively;
$\det(\bX)$ denotes the determinant of $\bX$;
$\Diag(x_1,\ldots,x_n)$ denotes a diagonal matrix with the $i$th diagonal element given by $x_i$;
$\bzero$ and $\bone$ denote all-zero and all-one vectors of appropriate sizes, respectively;
$\bx \geq \bzero$ means that $\bx$ is element-wise non-negative,
and similarly $\bX \geq \bzero$ means that $\bX$ is element-wise non-negative;
$\| \cdot \|$ denotes the Euclidean norm for both vectors and matrices;
$\conv(\bA)= \{ \by = \bA \bx \mid \bx \geq 0, \bone^\top \bx = 1 \}$ denotes the convex hull of the columns of $\bA$;
$ p(\bx;\bm \theta)$ denotes the probability distribution of a random variable $ \bx $, with the distribution parameter given by $ \bm\theta $;
$ p(\bx,\by;\bm \theta)$ denotes the joint probability distribution of two random variables $ \bx $ and $ \by $, with distribution parameter $\bm \theta $;
$ p(\bx | \by;\bm\theta)$ denotes the probability distribution of $ \bx $ conditioned on $ \by $, with distribution parameter $ \bm \theta $;
$\Exp[ \cdot ]$ denotes the expectation.
More notations will be defined in appropriate places.


\section{Background}

\subsection{Problem Statement}
\label{sect:prob_stat}

The problem of interest, in its most basic form, is as follows.
We are given a collection of data points $\by_1,\ldots,\by_T \in \Rbb^M$.
We postulate that
\beq \label{eq:basic_model}
\by_t = \bA_0 \bs_t,
\eeq
where $\bA_0 \in \Rbb^{M \times N}$, with $M \geq N$;
$\bs_t$ is a latent (and thus unknown) variable.
The latent variables lie in the unit simplex, i.e., $\bs_t \geq \bzero, \bone^\top \bs_t = 1$.
The matrix $\bA_0$ is unknown.
The problem is to recover $\bA_0$ from $\by_1,\ldots,\by_T$.
Note that after recovering $\bA_0$, we can recover $\bs_t$ by solving the regression problem $\min_{\bs_t \geq \bzero,  \bone^\top \bs_t=1} \| \by_t - \bA_0 \bs_t \|^2$.
For convenience, the above problem of recovering $\bA_0$ from $\by_1,\ldots,\by_T$ will be called SCA in the sequel.

From a geometrical viewpoint, SCA is a problem of finding the vertices of a hidden simplex from a collection of data points that lie in that simplex.
To be specific, observe from \eqref{eq:basic_model} that $\by_t \in \conv(\bA_0)$;
or, in words, the data points lie in $\conv(\bA_0)$.
The set $\conv(\bA_0)$ is a simplex under the assumption of full-column rank $\bA_0$, and, by the definition of simplices, the vertices of $\conv(\bA_0)$ are the columns of $ \bA_0$.\footnote{We should recall that a set $ \setS \subseteq \Rbb^{m} $ is called a simplex if it takes the form $ \setS = \conv(\bA) $, where $ \bA =[\ba_1,\ldots,\ba_n] \in \Rbb^{m\times n} $ has $ \{\ba_1,\ldots,\ba_n\} $ being affinely independent. A simplex $ \conv(\bA) $ has the property that the set of vertices of $ \conv(\bA) $ is $ \{\ba_1,\ldots,\ba_n\} $.
	Also, it should be noted that if $ \bA $ has full column rank, then $ \{\ba_1,\ldots,\ba_n\} $ is affinely independent; the converse is not true.}
Hence, the $\by_t$'s are simplicially distributed data, and recovering $\bA_0$ is the same as finding the vertices.
Such viewpoint is commonly used in the context of hyperspectral unmixing; see, e.g., \cite{Jose12,Ma2014HU}.
From a statistical viewpoint, SCA is reminiscent of latent factor analyses such as independent component analysis (ICA).
Specifically they share the common goal of exploiting the underlying natures of the latent variables, which are based upon further postulates on the statistics of the $\bs_t$'s, to recover $\bA_0$.
Note that unit-simplex distributed $\bs_t$'s do not have element-wise independent $\bs_t$'s, the latter being the key postulate of ICA.


An important application of SCA is hyperspectral unmixing (HU) in remote sensing \cite{Jose12,Ma2014HU}.
In fact, HU has provided strong motivations for researchers to study SCA, and one can argue that HU is central to the developments of SCA.
A concise problem statement of HU is as follows.
We are given a hyperspectral image taken from a scene.
The image is represented by $\by_1,\ldots,\by_T$,
where each $\by_t \in \Rbb^M$ is a collection of reflectance measurements over a number of $M$ (over a hundred) fine-resolution spectral bands at a particular pixel.
Under some assumptions we may postulate that $\by_t$ follows the SCA model \eqref{eq:basic_model} \cite{Jose12}.
In particular, each column of $\bA_0$ describes the spectral response of a distinct material (or endmember), and
each $\bs_t$ describes the proportional distribution (or abundance)
of the various materials at pixel $t$.
The problem of HU is to identify the unknown materials and how they compose the scene, specifically, by uncovering the materials' spectral responses and the proportional  distributions from the image.
The problem is, in essence, SCA. The reader is refered to \cite{Jose12,dobigeon2009joint,fu2016robust,Ma2014HU,Nascimento2005,nascimento2012hyperspectral,nascimento2009learning,PRISM2021} for further details of HU.

SCA has strong connections with non-negative matrix factorization (NMF).
To describe, consider an NMF data model $\bz_t = \bB \bc_t$ for $t=1,\ldots,T$,
where $\bB \geq \bzero$ and $\bc_t \geq \bzero$ for all $t$.
Note that $\bc_t$ may not satisfy $\bone^\top \bc_t = 1$.
Consider normalizing the data points $\bz_t$'s by $\by_t = \bz_t/( \bone^\top \bz_t )$.
One can show that
\[
\by_t  = \sum_{i=1}^N \underbrace{\frac{\bb_i}{ \bone^\top \bb_{i} }}_{:= {\ba}_{i,0}} \underbrace{\frac{ \bone^\top \bb_{i} c_{i,t} }{ \sum_{j=1}^N \bone^\top \bb_{j} c_{j,t} }}_{:= {s}_{i,t}} = {\bA}_0 {\bs}_t,
\]
where $\bb_i$ and $\ba_{i,0}$ denote the $i$th column of $\bB$ and $\bA_0$, respectively, and the above defined
${\bs}_t$ is seen to satisfy ${\bs}_t \geq \bzero$ and $\bone^\top {\bs}_t = 1$; see \cite{fu2019nonnegative,gillis2021nmf} and the references therein.
Thus, NMF can be cast as an SCA problem by the above normalization process.	It is worth noting that the application of SCA to NMF does not exploit the non-negativity of $\bA_0$ in general;
rather, it focuses on leveraging the structures of the unit-simplex-distributed $\bs_t$'s to recover $\bA_0$.
The reader is referred to \cite{fu2019nonnegative,gillis2021nmf} for details.

\subsection{Simplex Volume Minimization and SISAL}

There are various ways to tackle SCA, and, among them, simplex volume minimization (SVMin) stands as a powerful approach.
SVMin is built on the geometrical intuition that, if we can find a simplex that circumscribes all the data points and yields the minimum volume, that simplex is expected to be the ground-truth simplex $\conv(\bA_0)$;
see the literature \cite{Jose12,Ma2014HU,fu2019nonnegative,gillis2021nmf} for more inspirations.
The problem of finding the minimum-volume data circumscribing simplex can be formulated as
\beq \label{eq:svmin}
\begin{aligned}
	\min_{\bA \in \Rbb^{M \times N}} & ~ \svol(\bA):=  (N-1)! \cdot ( \det(\bbA^\top \bbA) )^{1/2} \\
	{\rm s.t.} & ~ \by_t \in \conv(\bA), \quad t=1,\ldots,T,
\end{aligned}
\eeq
where $\svol(\bA)$ is the volume of the simplex $\conv(\bA)$ \cite{Gritzmann1995} (we assume that every feasible point $\bA$ of \eqref{eq:svmin} has full column rank);
$\bbA= [~ \ba_1 - \ba_N, \ldots, \ba_{N-1} - \ba_N ~]$, with $\ba_i$ being the $i$th column of $\bA$.
Recent studies have revealed that SVMin is more than an intuition.
It is shown that, under some technical conditions which should hold for sufficiently well-spread $\bs_t$'s, the optimal solution to the SVMin problem~\eqref{eq:svmin} is the ground truth $\bA_0$ or its column permutation \cite{lin2015identifiability,Fu2015,fu2016robust}.
In other words, SVMin is equipped with provable recovery guarantees.

SISAL~\cite{Dias2009} is arguably the most popular algorithm for SVMin.
Here we shed light onto how SVMin is formulated in SISAL.
Bioucas-Dias, the author of SISAL, derived the SISAL formulation in an intuitively powerful way.
In particular, he focused on rewriting SVMin to a form that is algorithmically friendly to handle.
Assume $M=N$; this is not a problem since we can apply dimensionality reduction to project the data points to a lower dimensional space \cite{Jose12,Ma2014HU}.
SISAL starts with the following variation of writing the SVMin problem
\beq \label{eq:svmin_sisal}
\begin{aligned}
	\min_{\bA \in \Rbb^{N \times N}, \bS \in \Rbb^{N \times T}} & ~ | \det(\bA) | \\
	{\rm s.t.} & ~ \bY = \bA \bS, ~ \bS \geq \bzero, ~ \bS^\top \bone = \bone,
\end{aligned}
\eeq
where $\bY = [~ \by_1, \ldots, \by_T ~]$.
In particular the above problem replaces the simplex volume $ \svol(\bA) \propto ( \det(\bbA^\top \bbA) )^{1/2} $ in problem \eqref{eq:svmin} with $ | \det(\bA) | $---which is easier to work with.
The first key idea leading to SISAL is to perform a transformation
\[
\bB = \bA^{-1},
\]
for which we assume that every feasible point $\bA$ of problem \eqref{eq:svmin_sisal} is invertible.
By $\bY = \bA\bS \Longleftrightarrow \bB \bY = \bS$, we can transform problem \eqref{eq:svmin_sisal} to
\beq \label{eq:svmin_sisal2}
\begin{aligned}
	\min_{\bB \in \Rbb^{N \times N}} & ~ 1/| \det(\bB) | \\
	{\rm s.t.} & ~ \bB \bY  \geq \bzero, ~ \bY^\top \bB^\top \bone = \bone.
\end{aligned}
\eeq
The transformed problem above is a non-convex optimization problem with convex constraints, and in this regard we should note that the constraint $ \bY = \bA\bS $ in the SVMin problem \eqref{eq:svmin_sisal} is non-convex.
The second idea, which looks minor but will be relevant to a key aspect later, is to assume that
\beq \label{eq:equality_trick}
\bY^\top \bB^\top \bone = \bone \qquad \Longleftrightarrow  \qquad
\bB^\top \bone = (\bY^\top)^\dag  \bone.
\eeq
Note that \eqref{eq:equality_trick} is true for ``$\Longrightarrow$'',
but \eqref{eq:equality_trick} is not necessarily true for ``$\Longleftarrow$'' when we are given an arbitrary $\bY$.
Applying \eqref{eq:equality_trick}, we rewrite problem \eqref{eq:svmin_sisal2} as
\beq \label{eq:svmin_sisal3}
\begin{aligned}
	\min_{\bB \in \Rbb^{N \times N}} & ~ 1/| \det(\bB) | \\
	{\rm s.t.} & ~ \bB \bY  \geq \bzero, ~ \bB^\top \bone = (\bY^\top)^\dag  \bone.
\end{aligned}
\eeq
The constraint $\bB \bY  \geq \bzero$, albeit convex, is a number of $NT$ linear inequalities.
These linear inequalities are unstructured, meaning that there is no special structure that we can utilize to handle the inequalities efficiently.
When $T$ is large, which is often the case in practice, forcing the numerous linear inequalities to hold can be a computational challenge.
The third idea, which is a compromise, is to approximate the constraint $\bB \bY \geq \bzero$ by soft constraints.
This gives rise to the final formulation of SISAL:

\medskip
\begin{center}
	\noindent \fbox{\parbox{0.97\linewidth}{
			{\bf Formulation 1, SISAL Formulation by Bioucas-Dias \cite{Dias2009}:}
			\beq
			\begin{aligned}
				\min_{\bB \in \Rbb^{N \times N}} & ~ - \log ( | \det(\bB) | ) + \lambda \sum_{t=1}^T \sum_{i=1}^N {\rm hinge}( \bb_i^\top \by_t ) \\
				{\rm s.t.} & ~  \bB^\top \bone = (\bY^\top)^\dag  \bone,
			\end{aligned}
			\nonumber
			\eeq
			where ${\rm hinge}(x) = \max\{-x,0\}$ is a hinge function, and it serves as a penalty function for non-negative $x$;
			$\bb_i$ denotes the $i$th row of $\bB$;
			$\lambda > 0$ is a pre-selected penalty parameter;
			recall $\bB= \bA^{-1}$.
	}}
\end{center}
\medskip

Our description of the formulation of SISAL is complete.
Let us summarize the ideas that led to the SISAL formulation:
\begin{itemize}
	\item[i)] use the SVMin formulation \eqref{eq:svmin_sisal}, which considers $ M = N $ and replaces the simplex volume $ \svol(\bA) $ in \eqref{eq:svmin} with $ |\det(\bA) | $;
	\item[ii)] apply the variable transformation $ \bB = \bA^{-1} $;
	\item[iii)] assume that the equivalence in \eqref{eq:equality_trick} is true;
	\item[iv)] apply the soft constraint approximations, replacing the constraints $ \bB\bY \geq \bzero $ with a penalty function $ \lambda\sum_{t=1}^T \sum_{i=1}^N {\rm hinge}( \bb_i^\top \by_t ) $ in the objective function.
\end{itemize}

All these operations aim at simplifying the problem for efficient optimization. Interestingly it is recently shown that, except for operation iv), and under appropriate model assumptions,
all the above operations lead us to the same problem as the basic SVMin formulation in \eqref{eq:svmin}.

\begin{Prop}[{\cite{Ma2021}}] Suppose that the data points exactly follow the  data model $\by_t = \bA_0 \bs_t$, with $M = N$;
	that $\bA_0$ has full column rank;
	and that $\bS = [~ \bs_1,\ldots,\bs_T ~]$ has full row rank.
	Then, the SVMin problem \eqref{eq:svmin} is equivalent to problem \eqref{eq:svmin_sisal3}.
	Particularly, given any feasible point $\bA$ of problem \eqref{eq:svmin},
	(a) $\bA$ is invertible; (b) the both sides of the implications of \eqref{eq:equality_trick} are true; (c) it holds that $\svol(\bA) = C \cdot |\det(\bA)|$ for some constant $C$.
\end{Prop}

\subsection{Why is SISAL Successful?}

There are two reasons for the success of SISAL.
The first is with computational efficiency.
Bioucas-Dias built a specialized algorithm for Formulation 1, which is a combination of successive convex approximation and the variable splitting augmented Lagrangian method.
The result is a computationally efficient algorithm that scales well with the data size $T$, particularly compared to other SVMin algorithms that deal with the hard constraint $\bB \bY \geq \bzero$.
The second is with noise robustness.
The reader may have noticed that the SISAL formulation was derived under a data model that postulates that every data point is perfectly drawn from $\by_t = \bA_0 \bs_t$---with no noise.
As it turns out, the key success of SISAL lies in the noisy case.
The soft constraint approximation, which was at first introduced to avoid the hard constraint $\bB \bY \geq \bzero$, provides SISAL with resilience to noise effects.
It was noticed that SISAL can be robust to outlying data points, while SVMin algorithms that faithfully implement the hard constraint $\bB \bY \geq \bzero$ may not.
This gives SISAL a significant advantage in practice.

SISAL does have a weakness.
It is not clear how the penalty parameter $\lambda$ should be chosen, and usually it is manually tuned.


\section{SISAL as Probabilistic SCA, and Beyond}
\label{sect:prism}

Intriguingly, we can provide an explanation of why SISAL works in the noisy case.
The idea is to build a connection between SISAL and a probabilistic SCA framework,
and this is the focus of this section.

\subsection{Probabilistic SCA}

To put into context, consider a noisy data model
\beq \label{eq:prob_model}
\by_t = \bA_0 \bs_t + \bv_t, \quad t=1,\ldots,T,
\eeq
where $\bv_t$ is noise. The model is accompanied with the following assumptions:
\begin{itemize}
	\item[i)] $\bA_0$ is square and invertible;
	\item[ii)] every $\bs_t$ is uniformly distributed on the unit simplex; or, equivalently, every $\bs_t$ follows a Dirichlet distribution with concentration parameter $\bone$;
	\item[iii)] every $\bv_t$ is Gaussian distributed with mean zero and covariance $\sigma^2 \bI$;
	\item[iv)] the $\bs_t$'s are independent and identically distributed (i.i.d.),
	the $\bv_t$'s are i.i.d., and the $\bs_t$'s are independent of the $\bv_t$'s.
\end{itemize}
Our point of departure is the maximum-likelihood (ML) estimator
\beq \label{eq:mle}
\begin{aligned}
	\hat{\bA} \in \arg \max_{\bA \in \Rbb^{N \times N}} & ~ \frac{1}{T} \sum_{t=1}^T \log p(\by_t;\bA) \\
	{\rm s.t.} &  ~ \text{$\bA$ is invertible},
\end{aligned}
\eeq
where $p(\by;\bA)$ is the probability distribution of a data point $\by$ parameterized by $\bA$, which will be specified shortly.
The ML estimator \eqref{eq:mle} has been shown to possess a desirable identifiability characteristic \cite{PRISM2021}.
In addition, ML estimation is deemed a principled and powerful approach for estimating $\bA_0$ in the noisy case,
and the same type of ML estimation is also seen in probabilistic forms of principal component analysis (PCA) and ICA \cite{tipping1999probabilistic,pham1997blind,attias1999independent,khemakhem2020variational}.

\subsection{Approximating the Likelihood}

The expression of $p(\by;\bA)$ and how we handle it hold the first key of connecting SISAL and the ML estimator.
To derive $p(\by;\bA)$, let $p(\by,\bs;\bA)$ be the joint distribution of a data point $\by$ and its associated latent variable $\bs$ (parameterized by $\bA$).
From the model in \eqref{eq:prob_model} and its accompanying assumptions,
$p(\by,\bs;\bA)$ is given by
\begin{align}
p(\by,\bs;\bA) &  = p(\by|\bs;\bA) p(\bs), \\
p(\by|\bs;\bA) &  = \setN(\by; \bA\bs, \sigma^2 \bI),  \\
p(\bs)         &  = (N-1)! \cdot \indfn{\Delta}(\bs), \quad \Delta = \{ \bs \in \Rbb_{++}^N \mid \bone^\top \bs = 1 \},
\end{align}
where $p(\bs)$ is the latent prior; $p(\by|\bs;\bA)$ is the distribution of $\by$ conditioned on $\bs$ (and parameterized by $\bA$);
$\setN(\bx; \bmu, \bSig )$ denotes a real-valued multivariate Gaussian distribution function with mean $\bmu$ and covariance $\bSig$;
\[
\indfn{\setX}(\bx) = \left\{ \begin{array}{ll}
0 & \text{if  $\bx \notin  \setX$} \\
1 & \text{if  $\bx \in  \setX$}
\end{array}.  \right.
\]
The distribution $p(\by;\bA)$ is the marginalization of $p(\by,\bs;\bA)$ over $\bs$:
\beq \label{eq:likeli1}
p(\by;\bA) = \int p(\by,\bs;\bA) {\rm d}\mu(\bs),
\eeq
where $\mu$ is the Lebesgue measure on $\{ \bs \in \Rbb^N \mid \bone^\top \bs = 1 \}$.
At first sight, and by intuition, one may be tempted to further write \eqref{eq:likeli1} as
\beq \label{eq:likeli2}
p(\by;\bA) = \int_{\Rbb^N} p(\by,\bs;\bA) {\rm d}\bs.
\eeq
But the correct way should be
\beq
p(\by;\bA) = \int_{\Rbb^{N-1}} p(\by,(\bs_{1:N-1},1- \bone^\top \bs_{1:N-1});\bA) {\rm d}\bs_{1:N-1},
\nonumber
\eeq
where $\bs_{1:N-1}= (s_1,\ldots,s_{N-1})$, and
we use the relation $\bone^\top \bs = 1$ to explicitly represent $s_N$ by $s_N = 1- \bone^\top \bs_{1:N-1}$.
Simply speaking, \eqref{eq:likeli2} does not consider the mathematical caveat that $\indfn{\Delta}(\bs)$ is not measurable on $\Rbb^N$.
There is however a simple trick to get around this caveat and thereby allow us to use \eqref{eq:likeli2} (which is simpler), as we will study later.

The function in \eqref{eq:likeli1} requires us to solve an integral.
Unfortunately, that integral is intractable in general.
To be more precise, we do not know if there exists a simple analytical expression or a computationally efficient method to solve the integral, given an arbitrary instance of $\by,\bA,N$.
As with many scientific and engineering studies, we pursue approximations and heuristics.
Firstly, we adopt a quasi latent prior
\beq \label{eq:quasi_latentprior}
p(\bs) \simeq C \cdot \indfn{\hat{\Delta}}(\bs), \quad \hat{\Delta} = \{ \bs \in \Rbb_{++}^N \mid  | \bone^\top \bs - 1 | < \delta/2 \},
\eeq
where $\delta > 0$ is given and is small; $C$ is a normalizing constant.
Clearly, \eqref{eq:quasi_latentprior} should closely approximate the true latent prior when $\delta$ is very small.
Since the quasi latent prior \eqref{eq:quasi_latentprior} is measurable on $\Rbb^N$, we can use the expression \eqref{eq:likeli2} and write
\beq \label{eq:likeli_derive1}
p(\by;\bA) \simeq C \int_{\Rbb^N} \setN(\by;\bA\bs,\sigma^2 \bI)  \indfn{\hat{\Delta}}(\bs) {\rm d}\bs.
\eeq
Let $\bB = \bA^{-1}$. By the change of variable $\bx = \bA \bs$, \eqref{eq:likeli_derive1} can be rewritten as
\begin{align}
p(\by;\bA) & \simeq C | \det(\bB) | \int_{\Rbb^N} \setN(\by;\bx,\sigma^2 \bI)  \indfn{\hat{\Delta}}(\bB \bx) {\rm d}\bx
\nonumber \\
&
=  C | \det(\bB) | \int_{\Rbb^N} \setN(\bx;\by,\sigma^2 \bI)  \indfn{\hat{\Delta}}(\bB \bx) {\rm d}\bx.
\label{eq:likeli_derive2}
\end{align}
By another change of variable $ \bv = \bx -\by $, we can further rewrite \eqref{eq:likeli_derive2} as
\begin{align}
p(\by;\bA) & \simeq C | \det(\bB) | \int_{\Rbb^N} \setN(\bv;\bzero,\sigma^2 \bI)  \indfn{\hat{\Delta}}(\bB(\by+\bv)) {\rm d}\bv \nonumber \\
& = C | \det(\bB) | \cdot {\rm Prob}( \bB(\by+\bv) \in \hat{\Delta} ),
\label{eq:likeli_derive2_extra}
\end{align}
where $\bv \sim \setN(\bzero,\sigma^2 \bI)$.
By noting the definition of $ \hat{\Delta} $ in \eqref{eq:quasi_latentprior}, the probability term in \eqref{eq:likeli_derive2_extra} can be expressed as
\beq \label{eq:likeli_derive3}
{\rm Prob}( \bB(\by+\bv) \in \hat{\Delta} ) = {\rm Prob}\left( \bb_1^\top (\by + \bv)>0,\ldots,\bb_N^\top (\by + \bv)>0, | \bone^\top \bB(\by + \bv) - 1 | < \delta/2 \right),
\eeq
where $\bb_i$ denotes the $i$th row of $\bB$. For convenience, let
\begin{subequations}
	\begin{align}
	\setE_i & = \{ \bb_i^\top (\by + \bv) > 0 \}, \quad i=1,\ldots,N,
	\label{eq:E1}
	\\
	\setE_{N+1} & = \{ | \bone^\top \bB(\by + \bv) - 1 | < \delta/2 \},
	\label{eq:E2}
	\end{align}
\end{subequations}
and write
\[
{\rm Prob}\left(  \bB(\by+\bv) \in \hat{\Delta} \right) = {\rm Prob}\left(  \cap_{i=1}^{N+1} \setE_i \right).
\]
The following heuristic is very crucial.
\begin{Heuristic}
	Approximate \eqref{eq:likeli_derive3} by
	\[
	{\rm Prob}\left(  \cap_{i=1}^{N+1} \setE_i \right) \approx \prod_{i=1}^{N+1} {\rm Prob}( \setE_i ).
	\]
\end{Heuristic}
We will discuss how to make sense of Heuristic~1 in the next subsection.
One can show from \eqref{eq:E1} that
\[
{\rm Prob}( \setE_i ) = \Phi\left(  \frac{\bb_i^\top \by}{\sigma \| \bb_i \|} \right),
\quad i=1,\ldots,N,
\]
where $\Phi(x) = \frac{1}{\sqrt{2\pi}} \int_{-\infty}^x e^{-z^2/2} {\rm d}z$; the idea is that, for $ \bv \sim \setN(\bzero,\sigma^2 \bI) $, we have $ \bb_i^\top (\by + \bv) \sim \setN(\bb_i^\top \by,\sigma^2 \|\bb_i\|^2) $.
Also, we see from \eqref{eq:E2} that
\[
{\rm Prob}( \setE_{N+1} ) = \int_{-\delta/2}^{\delta/2}\setN( \eta; \bone^\top \bB \by -1, \sigma^2 \| \bB^\top \bone \|^2 )d\eta \simeq \delta \cdot \setN( 0; \bone^\top \bB \by -1, \sigma^2 \| \bB^\top \bone \|^2 )
\]
for a very small $ \delta $; again, the idea is that, for $ \bv \sim \setN(\bzero,\sigma^2 \bI) $, we have $ \bone^\top \bB(\by + \bv) - 1 \sim \setN(\bone^\top \bB \by -1, \sigma^2 \| \bB^\top \bone \|^2 ) $.
Putting the components together, we obtain an approximate expression of $p(\by;\bA)$ as follows
\beq \label{eq:approx_likeli}
p(\by;\bA) \approx \delta C |\det(\bB)| \cdot \left( \prod_{i=1}^N  \Phi\left(  \frac{\bb_i^\top \by}{\sigma \| \bb_i \|} \right) \right) \cdot  \setN( 0; \bone^\top \bB \by -1, \sigma^2 \| \bB^\top \bone \|^2 ).
\eeq

\subsection{Insights Revealed and Discussion}

Allow us to pause a moment to examine how the ML problem looks like under the likelihood approximation derived in the preceding subsection.
By applying \eqref{eq:approx_likeli} to the ML problem \eqref{eq:mle}, the following formulation can be shown.

\medskip

\begin{center}
	\noindent \fbox{\parbox{0.97\linewidth}{
			{\bf Formulation 2, An Approximate Formulation of the ML Problem \eqref{eq:mle}, Principally by Heuristic~1:}
			\beq
			\min_{\bB \in \Rbb^{N \times N}}  ~ - \log ( | \det(\bB) | ) + g(\bB) -  \frac{1}{T} \sum_{t=1}^T \sum_{i=1}^N \log \Phi \left( \frac{ \bb_i^\top \by_t}{ \sigma \| \bb_i \| } \right),
			\nonumber
			\eeq
			where we recall $\Phi(x) = \frac{1}{\sqrt{2\pi}} \int_{-\infty}^x e^{-z^2/2} {\rm d}z$;
			\[
			g(\bB) = \log(\| \bB^\top \bone \|) + \frac{\| \bY^\top \bB^\top \bone - \bone \|^2}{2 \sigma^2 T \| \bB^\top \bone \|^2 }.
			\]
	}}
\end{center}

\medskip

As a minor point of note for Formulation 2, we do not explicitly write down the constraint of invertible $\bB$, which comes from the constraint of invertible $\bA$ in the ML problem \eqref{eq:mle}.
This is because $-\log | \det(\bB) | = +\infty$ for non-invertible matrices, which means that the invertible matrix constraint is already taken care of.

\begin{figure}[hbt]
	\centering \includegraphics[width=0.35\linewidth]{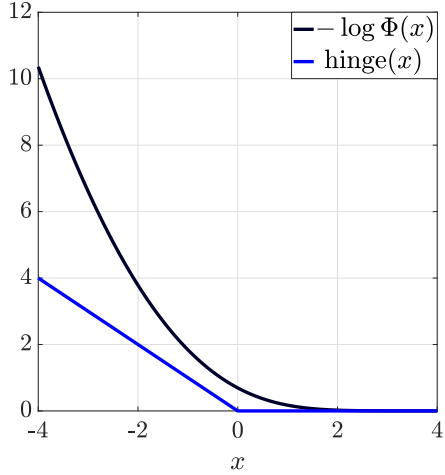}
	\caption{Comparison of $-\log \Phi(x)$ and the hinge function.}
	\label{fig:penaltyfunctionhinge}
\end{figure}

Let us compare Formulation 2 and the SISAL formulation (Formulation 1).
We see that both have penalty terms related to negative $\bb_i^\top \by_t$.
To better illustrate, Fig.~\ref{fig:penaltyfunctionhinge} plots $-\log \Phi(x)$ and the hinge function.
It is observed that  $-\log \Phi(x)$ is monotone decreasing, and it gives stronger outputs as $x$ is more negative.
Hence we may see $-\log \Phi(x)$ as a penalty function for negative $x$, serving a similar aim as the hinge function.
Moreover, the constraint $\bB^\top \bone = (\bY^\top )^\dag \bone$ in the SISAL formulation, which comes from $\bY^\top \bB^\top \bone = \bone$, is seen to bear some resemblance to the penalty function $g$ in Formulation 2.
In the next subsection, we will put forth another element that will bring Formulation~2 even closer to the SISAL formulation.
Some discussions are as follows.

\begin{Remark}
	Some related work should be mentioned. In \cite{PRISM2021}, we derived an approximate ML formulation similar to Formulation 2. We applied an approximation similar to Heuristic~1, but we did not use the quasi latent prior in \eqref{eq:quasi_latentprior}.
	As a result, our previous approximate ML formulation is still not as similar to SISAL as Formulation 2.
\end{Remark}

\begin{Remark}
	We return to the question of how we can make sense of Heuristic 1. Here is our intuition: By the probability result ${\rm Prob}\left(  \cap_{i=1}^{N+1} \setE_i \right) \leq {\rm Prob}(\setE_i)$ for any $i$, we have\[
	{\rm Prob}\left(  \cap_{i=1}^{N+1} \setE_i \right) \leq \left( \prod_{i=1}^{N+1} {\rm Prob}( \setE_i ) \right)^{1/(N+1)}.
	\]
	From the above inequality, we can show that
	\beq \label{eq:remark_reasoning}
	-\frac{1}{T} \sum_{t=1}^T \log p(\by;\bA) \geq - \log ( | \det(\bB) | ) + \frac{1}{N+1} \left[  g(\bB) -  \frac{1}{T} \sum_{t=1}^T \sum_{i=1}^N  \log \Phi \left( \frac{ \bb_i^\top \by_t}{ \sigma \| \bb_i \| } \right) \right],
	\eeq
	which is a lower-bound approximation and sounds better in terms of being equipped with a rationale.
	Empirically, we however found that \eqref{eq:remark_reasoning} tends to underestimate the negative log likelihood value $-\frac{1}{T} \sum_{t=1}^T \log p(\by;\bA)$ quite significantly.
	Instead, removing the scaling $1/(N+1)$ from \eqref{eq:remark_reasoning} would give better results.
	As future work, it would be interesting to analyze the approximation accuracy of Heuristic 1 or to study better approximations under the genre of Heuristic 1.
\end{Remark}

\subsection{Bringing SISAL and ML Closer}
\label{sect:prism_con}

We start with an assumption that does not seem to make sense at first.
Let
\[
\bp = \bA_0^{-\top} \bone,
\]
and {\em suppose} that we know $\bp$.
Consider the following modified ML problem
\beq \label{eq:mle_mod}
\begin{aligned}
	\max_{\bA \in \Rbb^{N \times N}} & ~  \frac{1}{T} \sum_{t=1}^T \log p(\by_t;\bA) \\
	{\rm s.t.} & ~ \bA^{-\top} \bone = \bp, \quad \text{$\bA$ is invertible},
\end{aligned}
\eeq
wherein we include our prior information of $\bp$ to better guide the estimation.
By applying the preceding likelihood approximation to problem~\eqref{eq:mle_mod} (or by adding the constraint $\bA^{-\top} \bone = \bp$ to Formulation 2), we have the following formulation.

\medskip

\begin{center}
	\noindent \fbox{\parbox{0.97\linewidth}{
			{\bf Formulation 3, An Approximate Formulation of the modified ML Problem \eqref{eq:mle_mod}, Principally by Heuristic~1:}
			\beq
			\begin{aligned}
				\min_{\bB \in \Rbb^{N \times N}}  & ~ - \log ( | \det(\bB) | ) -  \frac{1}{T} \sum_{t=1}^T \sum_{i=1}^N \log \Phi \left( \frac{ \bb_i^\top \by_t}{ \sigma \| \bb_i \| } \right) \\
				{\rm s.t.} & ~ \bB^\top \bone = \bp.
			\end{aligned}
			\nonumber
			\eeq
	}}
\end{center}

\medskip

Formulation 3 is very similar to the SISAL formulation (Formulation 1) if $\bp = (\bY^\top)^\dag \bone$.
In fact, we have this surprising result.
\begin{Fact}[\cite{Ma2021}] \label{fact:con1}
	Suppose that the data points $\by_t$'s follow the noiseless model $\by_t  = \bA_0 \bs_t$ (with $M= N$); that $\bA_0$ has full column rank; and that $\bS = [~ \bs_1,\ldots,\bs_T ~]$ has full row rank. Then,
	\[
	(\bY^\top)^\dag \bone = \bA_0^{-\top} \bone.
	\]
\end{Fact}
Fact~\ref{fact:con1} was shown in \cite{Ma2021}, and we shall not repeat the proof.
Rather, we are interested in its extension to the noisy case.
\begin{Fact} \label{fact:con2}
	Suppose that the data points $\by_t$'s follow the model in \eqref{eq:prob_model} and the accompanying assumptions. Let $\bmu_y = \Exp[ \by_t ]$ and $\bR_{yy} = \Exp[ \by_t \by_t^\top ]$ be the mean and correlation matrix of $ \by_t $, respectively. Then,
	\[
	( \bR_{yy} - \sigma^2 \bI )^{-1} \bmu_y = \bA_0^{-\top} \bone.
	\]
\end{Fact}

{\em Proof of Fact~\ref{fact:con2}:} \
Let $\bR_{ss} = \Exp[ \bs_t \bs_t^\top ]$, $\bmu_s  = \Exp [ \bs_t ]$. It can be verified that $\bR_{ss}$ is positive definite.
Also, from the data model \eqref{eq:prob_model}, we can show that
\[
\bR_{yy} = \bA_0 \bR_{ss} \bA_0^\top + \sigma^2 \bI,
\quad
\bmu_y = \bA_0 \bmu_s.
\]
It follows that
\[
( \bR_{yy} - \sigma^2 \bI )^{-1} \bmu_y = ( \bA_0 \bR_{ss} \bA_0^\top )^{-1} \bA_0 \bmu_s
= \bA_0^{-\top} \bR_{ss}^{-1} \bmu_s.
\]
It can be shown that $\bR_{ss}^{-1} \bmu_s = \bone$. Specifically,
\[
\bR_{ss} \bone = \Exp [ \bs_t \underbrace{\bs_t^\top \bone}_{=1} ] =  \Exp [ \bs_t ] = \bmu_s.
\]
The proof is complete.
Note that this result also applies to a more general case wherein $\bs_t$ follows a (and possibly non-uniform) $\Delta$-supported distribution with positive definite $\bR_{ss}$.
\hfill $\blacksquare$
\vspace{1em}

Fact~\ref{fact:con2} provides us with an implication that, in practice,
we can estimate $\bp$ by
\beq \label{eq:p_est}
\hat{\bp} = (\hat{\bR}_{yy} - \sigma^2 \bI)^{-1} \hat{\bmu}_y,
\quad \hat{\bR}_{yy} = \frac{1}{T} \sum_{t=1}^T \by_t \by_t^\top,
\quad
\hat{\bmu}_y = \frac{1}{T} \sum_{t=1}^T \by_t.
\eeq
Our final touch is to explain how the negative penalty terms in Formulation 3 and the SISAL formulation are related.
We start from the direction of Formulation 3.
Consider the following result.
\begin{Fact} {\bf (\cite{verdu1998multiuser}, \cite[footnote~1]{shao2019framework})} \label{fac:hinge1}
	It holds that $\Phi(x) \leq \frac{1}{2} e^{\sqrt{\frac{2}{\pi}} x }$.
	Also, as a direct consequence,
	\[
	- \log \Phi(x) \geq -\log \left( \max\left\{ \frac{1}{2} e^{\sqrt{\frac{2}{\pi}} x }, 1  \right\}  \right) = \max\left\{ \log(2) - \sqrt{\frac{2}{\pi}} x, 0 \right\}.
	\]
\end{Fact}
Using Fact~\ref{fac:hinge1}, the penalty terms of Formulation 3 can be approximated by
\begin{align}
- \log \Phi \left( \frac{ \bb_i^\top \by_t}{ \sigma \| \bb_i \| } \right) &
\geq  \max \left\{  \log(2) - \sqrt{\frac{2}{\pi}} \frac{ \bb_i^\top \by_t}{ \sigma \| \bb_i \| }, 0  \right\} \nonumber \\
& \geq \max \left\{  - \sqrt{\frac{2}{\pi}} \frac{ \bb_i^\top \by_t}{ \sigma \| \bb_i \| }, 0  \right\} \nonumber \\
& =  \sqrt{\frac{2}{\pi}} \frac{1}{\sigma \| \bb_i \|} {\rm hinge}(  \bb_i^\top \by_t ).
\end{align}
The normalizing term $\| \bb_i \|$ is hard to deal with.
By pretending as if $\| \bb_i \|$ were a constant, and by setting $\sqrt{\frac{2}{\pi}} \frac{1}{\sigma \| \bb_i \| T}  = \lambda$ for some pre-selected $\lambda  >0$,
we have
\beq
- \frac{1}{T} \log \Phi \left( \frac{ \bb_i^\top \by_t}{ \sigma \| \bb_i \| } \right)
\approx \lambda \cdot {\rm hinge}( \bb_i^\top \by_t ).
\label{eq:heuristic_hinge}
\eeq
Now, we are ready to draw our main conclusion:
{\em SISAL can be explained as an approximation of the ML estimator \eqref{eq:mle_mod}.}
In particular, the connection is made by applying Fact~\ref{fact:con1} and \eqref{eq:heuristic_hinge} to Formulation~3.

\subsection{A Hinge-Square Variant of SISAL}

\begin{figure}[hbt]
	\centering \includegraphics[width=.35\linewidth]{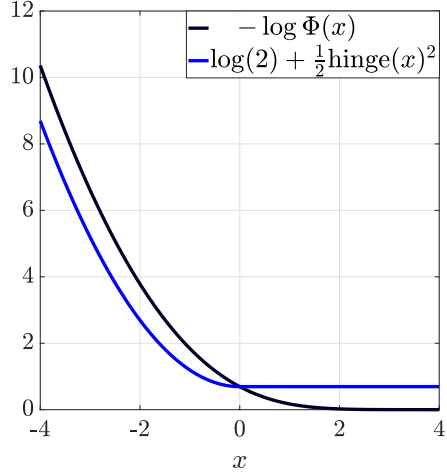}
	\caption{Comparison of $-\log \Phi(x)$ and a hinge-square based function.}
	\label{fig:penaltyfunctionhingesquare}
\end{figure}

The explanation of SISAL as an approximate ML estimator in the preceding subsection gives us a new insight, namely, that the hinge function serves as a surrogate of the penalty function $-\log \Phi(x)$ from the ML viewpoint.
In that regard, we can choose a different surrogate of $-\log \Phi(x)$.
From Fig.~\ref{fig:penaltyfunctionhinge} we see that, as $x$ becomes more negative, the hinge function is a poor approximation of $-\log \Phi(x)$.
Consider the following result.
\begin{Fact} {\bf (Chernoff bound; see, e.g.,  \cite{verdu1998multiuser})} \label{fac:hinge2}
	It holds that, for $x \leq 0$, $\Phi(x) \leq \frac{1}{2} e^{-x^2 /2 }$.
	Also, as a direct consequence, we may approximate
	\[
	-\log \Phi(x) \approx -\log \left(  \frac{1}{2} e^{- \max\{-x,0 \} ^2 /2 }  \right)
	= \log(2) + \frac{1}{2} {\rm hinge}(x)^2.	
	\]
\end{Fact}
Fig.~\ref{fig:penaltyfunctionhingesquare} compares the above surrogate and $-\log\Phi(x)$.
We see that this new surrogate approximates $-\log\Phi(x)$ better for negative $x$.
By approximating
\beq
- \frac{1}{T} \log \Phi \left( \frac{ \bb_i^\top \by_t}{ \sigma \| \bb_i \| } \right)
\approx \lambda \cdot {\rm hinge}( \bb_i^\top \by_t )^2 + {\rm constant},
\label{eq:heuristic_hinge2}
\eeq
as before, we have the following variant of SISAL.

\medskip
\begin{center}
	\noindent \fbox{\parbox{0.97\linewidth}{
			{\bf Formulation 4, H$^2$-SISAL;
				a Chernoff bound-based heuristic of the approximate ML problem in Formulation 3, or a  hinge-square variant of SISAL in Formulation 1:}
			\beq
			\begin{aligned}
				\min_{\bB \in \Rbb^{N \times N}} & ~ - \log ( | \det(\bB) | ) + \lambda \sum_{t=1}^T \sum_{i=1}^N {\rm hinge}( \bb_i^\top \by_t )^2 \\
				{\rm s.t.} & ~  \bB^\top \bone = \bp,
			\end{aligned}
			\nonumber
			\eeq
			where
			$\lambda > 0$ is a pre-selected penalty parameter.
	}}
\end{center}

Observe that the difference between Formulation 4 and the SISAL formulation (Formulation 1) is that the former puts a square on the hinge function.
From an optimization viewpoint,  this H$^2$-SISAL formulation has the advantage that the hinge-square penalty terms, as well as the whole objective function, are continuously differentiable.

\section{SISAL as an Algorithm, and More}

Having explored the formulation aspects with SISAL, we turn to the algorithmic aspects.
To facilitate our subsequent development, let us introduce some notations.
Let $f: \Rbb^n \rightarrow \Rbb \cup \{ + \infty \}$ be an extended real-valued function.
We denote
${\rm dom}\, f = \{ \bx \in \Rbb^n \mid f(\bx) < +\infty \}$ as the domain of $f$;
$\nabla f(\bx)$ as the gradient of $f$ (when $f$ is differentiable at $\bx$);
\[
{\rm prox}_f(\bx) \in \arg \min_{\bz \in \Rbb^n } \frac{1}{2} \| \bz - \bx \|^2 + f(\bx)
\]
as a proximal operator associated with $f$.
We also denote $\langle \cdot, \cdot \rangle$ as the inner product;
\[
\Pi_\setX(\bx) \in \arg \min_{\bz \in \setX} \| \bz - \bx \|^2
\]
as a projection of $\bx$ onto a closed set $\setX \subseteq \Rbb^n$;
\[
\mathbb{I}_\setX(\bx) = \left\{ \begin{array}{ll} +\infty & \text{if $\bx \notin \setX$} \\ 0 & \text{if $\bx \in \setX$} \end{array} \right.
\]
as the indicator function associated with $\setX$. Furthermore, we call $ f $ to have Lipschitz continuous gradient on $\setX $ if $ \nabla f $ is Lipschitz continuous on $\setX$; i.e., there exists $ \alpha > 0 $ such that $ \| \nabla f(\bx) - \nabla f(\by)\| \leq \alpha \| \bx - \by \| $ for all $ \bx, \by \in \setX $.

\subsection{The SISAL Algorithm}

To describe the algorithm used in SISAL, we start with describing the basic natures of the SISAL problem.
Recall from Formulation 1 the SISAL problem:
\beq \label{eq:sisal_prob_again}
\min_{\bB \in \Rbb^{N \times N}, \bB^\top \bone = \bp } ~ f(\bB) = \underbrace{ -\log | \det(\bB) | }_{:= f_0(\bB)} + \textstyle \lambda \sum_{t=1}^T \sum_{i=1}^N {\rm hinge}(\bb_i^\top \by_t ),
\eeq
where $\bp= (\bY^\top)^\dag \bone$.
The problem is non-convex and non-smooth:
the second term of $f$, which has the hinge function involved, is convex and non-differentiable;
$f_0$ is non-convex and continuously differentiable on its domain ${\rm dom}\, f_0$;
${\rm dom}\, f_0$ is the set of all invertible matrices on $\Rbb^{N \times N}$;
$f_0$ does {\em not} have Lipschitz continuous gradient on ${\rm dom}\, f_0$.
If one wants to find an off-the-shelf optimization method that offers some form of guarantee of finding a stationary point of problem \eqref{eq:sisal_prob_again}, that will not be immediately obvious.
The non-triviality comes in two ways:
\begin{enumerate}
	\item Implementation:
	One can actually apply an off-the-shelf method from the recent advances of optimization, particularly, first-order optimization.
	Take the proximal gradient method as an example.
	One needs to choose the step size, which is typically guided by the Lipschitz constant of $\nabla f_0$.
	The absence of Lipschitz continuous $ \nabla f_0$ in our problem necessitates a different strategy to deal with the problem.
	Also, the problem domain, the set of all invertible matrices, is non-standard at first sight.
	
	\item Theory: The Lipschitz continuity of $\nabla f_0$ is needed in most convergence proofs.
	Again, we do not have Lipschitz continuous $\nabla f_0$.
\end{enumerate}

Back to 2009,
Bioucas-Dias dealt with the problem by successive convex approximation.
The ideas are to form a quadratic approximation of $f_0$ at a given point $\tilde{\bB} \in {\rm dom}\, f_0$
\[
f(\bB) \approx f_0(\tilde{\bB}) + \langle \nabla f_0( \tilde{\bB} ), \bB - \tilde{\bB} \rangle + \frac{\mu}{2} \| \bB - \tilde{\bB} \|^2 := g_\mu(\bB, \tilde{\bB}),
\]
for some $\mu > 0$; and to solve, iteratively,
\beq \label{eq:sisal_algo1}
\bB^{k+1} = \arg \min_{\bB \in \Rbb^{N \times N}, \bB^\top \bone = \bp } g_{\mu_k}(\bB,\bB^k) + \textstyle \lambda \sum_{t=1}^T \sum_{i=1}^N {\rm hinge}(\bb_i^\top \by_t ),
\quad k=0,1,2,\cdots
\eeq
for some $ \mu_k > 0$ for all $k$.
The problems encountered in \eqref{eq:sisal_algo1} are convex (in fact, strictly convex).
Bioucas-Dias solved these problems by a variable splitting augmented Lagrangian algorithm, which is now more popularly known as the alternating direction method of multipliers (ADMM).
That ADMM algorithm exploits the problem structure of \eqref{eq:sisal_algo1} and is computationally efficient.
But \eqref{eq:sisal_algo1} has a caveat:
depending on how $\mu_k$ is chosen, a new iterate $\bB^{k+1}$ may not be invertible;
and when that happens, the successive convex optimization in \eqref{eq:sisal_algo1} will crash.

{\begin{algorithm}[hbt]
		\caption{SISAL by Bioucas-Dias \cite{Dias2009}, successive convex optimization for Formulation 1} \label{alg:sisal}
		\begin{algorithmic}[1]
			\STATE \textbf{given:} an invertible starting point $\bB^0$ and a constant $\mu > 0$
			\STATE $k= 0$
			\STATE {\bf repeat}
			\STATE \hspace{1em} $\displaystyle \bar{\bB}^{k} = \arg \min_{\bB \in \Rbb^{N \times N}, \bB^\top \bone = \bp } g_{\mu}(\bB,\bB^k) + \textstyle \lambda \sum_{t=1}^T \sum_{i=1}^N {\rm hinge}(\bb_i^\top \by_t )$, by ADMM (see \cite{Dias2009})
			\STATE \hspace{1em} find a $\theta_k \in (0,1]$ such that $f(\bB^k + \theta_k (\bar{\bB}^k - \bB^k )) \leq f(\bB^k)$, by line search
			\STATE \hspace{1em} $\bB^{k+1} = \bB^k + \theta_k (\bar{\bB}^k - \bB^k )$
			\STATE \hspace{1em} $k= k+1$
			\STATE {\bf until} a stopping rule is satisfied
			\STATE {\bf output:} $\bB^k$
		\end{algorithmic}
\end{algorithm}}

Algorithm \ref{alg:sisal} is the actual form of the SISAL algorithm. Intuitively, we expect that there should exist a $\theta_k \in (0,1]$, no matter how small it may be, such that $\bB^{k+1} =  \bB^k + \theta_k (\bar{\bB}^k - \bB^k )$ remains invertible.
As mentioned, empirical studies suggest that SISAL works.
This leads to an intriguing, and previously unanswered, basic question:
Does Algorithm~\ref{alg:sisal} have any guarantee of finding a stationary point of problem \eqref{eq:sisal_prob_again}?

\subsection{Line Search-Based Proximal Gradient Method}
\label{sect:lsb_pg}

Our study found that the optimization framework by Bonettini {\em et al.} \cite{BLPP16} can be used to answer the question.
To put into context, consider a problem
\beq \label{eq:prob_basic}
\min_{\bx \in \Rbb^n } f(\bx) := f_0(\bx) + f_1(\bx),
\eeq
where
$f_0$ is continuously differentiable on its domain ${\rm dom}\, f_0$;
${\rm dom}\, f_0$ is open;
$f_1$ is convex, proper, lower semicontinuous, and bounded from below;
${\rm dom}\, f_1$ is closed and nonempty.
For this problem, a point $\bar{\bx} \in {\rm dom}\, f$ is called a stationary point of problem \eqref{eq:prob_basic}  if the directional derivative of $f$, defined as $f'(\bx;\bd) = \lim_{t \downarrow 0} (f(\bx + t \bd) - f(\bx) )/t$, satisfies $f'(\bar{\bx};\bd) \geq 0$ for all $\bd \in \Rbb^n$.
To describe the method, let
\[
h_{\mu}(\bz,\bx)  =  \langle \nabla f_0(\bx), \bz - \bx \rangle + \frac{\mu}{2} \| \bz - \bx \|^2 + f_1(\bz) - f_1(\bx), \qquad \mu > 0.
\]
Consider the following line search-based proximal gradient (LSB-PG) method:
given $\beta \in (0,1)$, $\bx^0 \in {\rm dom}\, f$, recursively compute
\begin{align}
\by^k & = \arg \min_{\bz \in \Rbb^n} h_{\mu_k}(\bz,\bx^k) = {\rm prox}_{\mu_k^{-1} f_1}( \bx^k - \mu_k^{-1} \nabla f_0(\bx^k) ), \quad \text{for some $\mu_k > 0$,} \label{eq:lsb_pg1} \\
\bx^{k+1} & = \bx^k + \theta_k ( \by^k - \bx^k ), \label{eq:lsb_pg2}
\end{align}
for $k=0,1,2,\cdots$, where $\theta_k \in (0,1]$ is chosen such that
\beq \label{eq:suff_dec}
f( \bx^k + \theta_k ( \by^k - \bx^k ) ) \leq
f( \bx^k ) + \beta \theta_k h_{\mu_k}( \by^k, \bx^k).
\eeq
To be precise, we use an Armijo line search rule to find $\theta_k$:
find the smallest non-negative integer $j$ such that
\beq \label{eq:lsb_pg4}
f( \bx^k + \delta^j ( \by^k - \bx^k ) ) \leq
f( \bx^k ) + \beta \delta^j h_{\mu_k}( \by^k, \bx^k).
\eeq
for some given $\delta \in (0,1)$, and then choose $\theta_k = \delta^j$.
It is worth noting that \eqref{eq:suff_dec} is a sufficient decrease condition with the objective value, since $h_{\mu_k}(\by^k,\bx^k) \leq 0$.
Also, the framework in \cite{BLPP16} is much more general than the LSB-PG, and here we reduce the framework to the above minimal form which is enough to answer our question.

The LSB-PG method is equipped with the following stationarity guarantee.

\begin{Prop}[a rephrased, simplified, version of Corollary~3.1 in \cite{BLPP16}]
	Consider problem \eqref{eq:prob_basic} and its associated LSB-PG method in \eqref{eq:lsb_pg1}--\eqref{eq:lsb_pg4}.
	Suppose ${\rm dom}\, f_0 \supseteq {\rm dom}\, f_1$.
	Also, assume that $\{ \mu_k \} \subset [ \mu_{\rm min}, \mu_{\rm max} ]$ for some $0 < \mu_{\rm min} \leq \mu_{\rm max} < +\infty$, and that $\{ \bx_k \}$ has a limit point.
	Then any limit point of $\{ \bx_k \}$ is a stationary point of problem \eqref{eq:prob_basic}.
\end{Prop}

As we will discuss in the next subsection, the application of the LSB-PG method to the SISAL problem does not have ${\rm dom}\, f_0 \supseteq {\rm dom}\, f_1$ satisfied.
This led us to rework the whole proof to see if the above assumption can be relaxed.
The answer, fortunately, is yes.

\begin{Corollary}
	The same stationarity result in Proposition 2 holds if we replace ${\rm dom}\, f_0 \supseteq {\rm dom}\, f_1$ by ${\rm dom}\, f_0 \cap {\rm dom}\, f_1 \neq \emptyset$.
	As a comment, the assumption of open ${\rm dom}\, f_0$ plays a crucial role.
\end{Corollary}

The proof of Corollary 1 is a meticulous re-examination of the whole proof of Corollary~3.1 in \cite{BLPP16}, including the proof of the theorems and propositions that precede it.
We shall omit the proof.
The following remark describes the unique aspect of proving Corollary 1, and the reader may choose to skip it and jump to the next subsection for the application of Corollary 1 to the SISAL problem.

\begin{Remark}
	We discuss the key proof differences of Proposition~2 and Corollary~1.
	In the proof, an important issue is to show that there exists a $\theta_k \in (0,1]$ such that the sufficient decrease condition \eqref{eq:suff_dec} holds.
	To achieve the latter, a prerequisite is to ensure $\bx^{k+1} \in {\rm dom}\, f_0$.
	One can readily see from \eqref{eq:lsb_pg1}--\eqref{eq:lsb_pg2} that $\by^k \in {\rm dom}\, f_1$, and then $\bx^{k+1} \in {\rm dom}\, f_1$ (due to the convexity of ${\rm dom}\, f_1$).
	For the case of ${\rm dom}\, f_0 \supseteq {\rm dom}\, f_1$, or Proposition 2, we automatically get $\bx^{k+1} \in {\rm dom}\, f_0$.
	For the case of ${\rm dom}\, f_0 \nsupseteq {\rm dom}\, f_1$, or Corollary 1, we need to leverage on the assumption of open ${\rm dom}\, f_0$.
	Since ${\rm dom}\, f_0$ is open, there exists $\epsilon_k > 0$ such that, for any $\bu \in \Rbb^n$ with  $\|  \bu \| \leq \epsilon$, we have $\bx^k + \bu \in {\rm dom}\, f_0$.
	This implies that there must exist a $\theta_k > 0$, no matter how small it is, such that $\bx^k + \theta_k (\by^k - \bx^k ) \in {\rm dom}\, f_0$.
	The above is the distinct part of the proof of Corollary~1.
\end{Remark}

\subsection{Stationarity Guarantee of SISAL}

Now we apply the framework in the preceding subsection to the SISAL problem.
Let
\begin{equation*}
\begin{split}
f_0(\bB) = &~- \log |\det(\bB) |,	\\
f_1(\bB) = &~\textstyle \lambda \sum_{t=1}^T \sum_{i=1}^N {\rm hinge}(\bb_i^\top \by_t ) + \mathbb{I}_\setB(\bB), \quad \setB =~ \{ \bB \in \Rbb^{N \times N} \mid \bB^\top \bone = \bp \},
\end{split}
\end{equation*}
and let $\mu_k = \mu$ for some pre-selected constant $\mu > 0$.
We observe that the SISAL algorithm in Algorithm~\ref{alg:sisal} is very similar to the LSB-PG method in \eqref{eq:lsb_pg1}--\eqref{eq:lsb_pg4}, with $\beta$ being nearly zero.
Or, more specifically, if we modify Algorithm~\ref{alg:sisal} by changing the line search in Step 5 to the Armijo rule in \eqref{eq:lsb_pg4}, the algorithm is, faithfully, an instance of the LSB-PG method.
To answer the question of stationarity guarantees, note that ${\rm dom}\, f_0$ is the set of all invertible matrices on $\Rbb^{N \times N}$, while ${\rm dom}\, f_1 = \setB$.
Clearly, we have ${\rm dom}\, f_0 \nsupseteq {\rm dom}\, f_1$, and Proposition 2 is not applicable.
Corollary 1 is applicable if  ${\rm dom}\, f_0$ is open.
In fact, it is known in topology that the set of invertible matrices is open.\footnote{For the reader's interest, here is a simple proof by matrix analysis.
	Let $\setS$ be the set of invertible matrices on $\Rbb^{N \times N}$.
	Let $\bX \in \setS$, and let $\sigma_1 \geq \cdots \geq \sigma_N > 0$ be its singular values.
	Let $\epsilon > 0$.
	Let $\bY$ be any matrix such that $\| \bX - \bY \| \leq \epsilon$, and let $d_1  \geq \cdots \geq d_N \geq 0$ be its singular values.
	By the singular value inequality $\| \bX - \bY \|^2 \geq \sum_{i=1}^N | \sigma_i - d_i |^2$, and letting $\epsilon = \sigma_N/2$, one can verify that $d_N \geq \sigma_N/2 > 0$.
}
Let us conclude. By Corollary 1, the SISAL algorithm, upon a minor modification with its line search rule, is equipped with a stationarity guarantee.

%
%
%
%

\subsection{Application to H$^2$-SISAL and Formulation 3}

It is exciting to point out that we can also use the LSB-PG method in Section \ref{sect:lsb_pg} to deal with the H$^2$-SISAL problem in Formulation 4.
Specifically we choose
\beq \label{eq:f0_h2_sisal}
f_0(\bB) = - \log ( | \det(\bB) | ) + \textstyle \lambda \sum_{t=1}^T \sum_{i=1}^N {\rm hinge}( \bb_i^\top \by_t )^2, \quad f_1(\bB) = \textstyle  \mathbb{I}_\setB(\bB);
\eeq
note that we put the (continuously differentiable) hinge-square penalty term to $f_0$, which is different compared to  SISAL.
The resulting LSB-PG method has the proximal operation \eqref{eq:lsb_pg1} reduced to
\[
\bar{\bB}^k = {\rm prox}_{\mu_k^{-1} f_1}( \bB^k - \mu_k^{-1} \nabla f_0(\bB^k) ) = \Pi_{\setB}( \bB^k - \mu_k^{-1} \nabla f_0(\bB^k) ),
\]
which has a simple closed form and is cheap to compute.
We should recall that the proximal operation in SISAL has no closed form and requires us to call a solver (ADMM).
We take advantage of the computational efficiency of the proximal operation by considering the following rule of choosing $\mu_k$:
find the smallest non-negative integer $j$ such that
\begin{subequations} \label{eq:backtrack}
	\begin{align}
	f(\bar{\bB}^{k,j}) & \leq f(\bB^k) + \beta h_{\nu c^j}( \bar{\bB}^{k,j}, \bB^k), \\
	\bar{\bB}^{k,j} & = \Pi_{\setB}( \bB^k - (\nu c^j)^{-1} \nabla f_0(\bB^k) ),
	\end{align}
\end{subequations}
for some given $\nu > 0, c > 1$, and then choose $\mu_k = \nu c^j$.
Consequently, the sufficient decrease condition \eqref{eq:suff_dec} will be satisfied for $\theta_k = 1$, and we can simply set $\theta_k = 1$, $\bB^{k+1} = \bar{\bB}^{k,j}$.
Note that this is a typical scheme in proximal gradient methods  (see, e.g., \cite{beck2017first}), and \eqref{eq:backtrack} is popularly called the backtracking line search.
We should also mention that the above LSB-PG scheme is identical to the projected gradient method, with a suitably chosen step size.
By Corollary 1,
this LSB-PG scheme is equipped with a stationarity guarantee under the assumption that the $\mu_k$'s found by the backtracking line search are bounded.

Our actual algorithm, shown in Algorithm~\ref{alg:h2sisal}, is an extrapolated variant of the above scheme.
\begin{algorithm}[hbt!]
	\caption{H$^2$-SISAL, an extrapolated proximal gradient scheme for Formulation 4} \label{alg:h2sisal}
	\begin{algorithmic}[1]
		\STATE \textbf{given:} an invertible starting point $\bB^0$; a constant $\beta \in (0,1)$; and an extrapolation sequence $\{ \alpha_k \}$, typically the FISTA sequence \cite{beck2017first}
		\STATE $k= 0$, $\bB^{-1} = \bB^0$
		\STATE {\bf repeat}
		\STATE \hspace{1em} $\bB_{\rm ex}^k = \bB^k + \alpha_k ( \bB^k - \bB^{k-1} )$
		\STATE \hspace{1em} $\bB^{k+1} = \Pi_{\setB}( \bB^k_{\rm ex} - \mu_k^{-1} \nabla f_0(\bB^k_{\rm ex}) )$, where $\mu_k$ is chosen such that $f(\bB^{k+1}) \leq f(\bB_{\rm ex}^k) + $\\
		\STATE \hspace{1em} $\beta h_{\mu_k}( \bB^{k+1}, \bB^k_{\rm ex})$, done by the backtracking line search \eqref{eq:backtrack}; $f_0$ is given in \eqref{eq:f0_h2_sisal}
		\STATE \hspace{1em} $k= k+1$
		\STATE {\bf until} a stopping rule is satisfied
		\STATE {\bf output:} $\bB^k$
	\end{algorithmic}
\end{algorithm}
Note that, by choosing $\alpha_k = 0$, Algorithm~\ref{alg:h2sisal} reduces to the previous LSB-PG scheme.
Our consideration is more from the practical side.
The LSB-PG framework does not cover the extrapolated variant, and hence it is not known if Algorithm~\ref{alg:h2sisal} is equipped with stationarity guarantees.
On the other hand, we want to leverage on the merits of extrapolation demonstrated in prior works.
It is known that, when $f_0$ is convex and has Lipschitz continuous gradient, the extrapolated proximal gradient method can lead to faster convergence rates than the proximal gradient method, both provably and empirically \cite{beck2009fast};
and that, when $f_0$ is non-convex and has Lipschitz continuous gradient,  the extrapolated proximal gradient method is shown to yield some stationarity guarantee \cite{ghadimi2016accelerated,xu2017globally}, and similar methods were empirically found to lead to faster convergence speeds in some applications \cite{xu2013block,shao2019framework,wu2020hybrid,fu2016robust}.
Our empirical experience with Algorithm~\ref{alg:h2sisal} is good in terms of runtime speed and stability.

We should further note that all the developments in this subsection apply to the approximate ML problem in Formulation 3; change
\[
f_0(\bB) = - \log ( | \det(\bB) | ) -  \frac{1}{T} \sum_{t=1}^T \sum_{i=1}^N \log \Phi \left( \frac{ \bb_i^\top \by_t}{ \sigma \| \bb_i \| } \right)
\]
(this $f_0$ can be shown to be continuously differentiable on the set of all invertible matrices).
Unfortunately, by our numerical experience, the adaptation of Algorithm~\ref{alg:h2sisal} (with or without extrapolation) to Formulation 3 is not promising:
its convergence tends to be slow; and numerical instability could happen, if not careful enough.
The culprit is most likely the normalizing terms $\| \bb_i \|$:
the term $1/\| \bb_i \|$ becomes very large for small $\| \bb_i \|$,
and the occurrence of such event can cause numerical instability.
These setbacks drove us to rethink our strategy for dealing with Formulation~3.

\section{Probabilistic SISAL via Inexact Block Coordinate Descent}

In this section we devise an algorithm for tackling the approximate ML problem in Formulation~3, with a focus on practicality and efficiency in our design.

\subsection{Reformulation and Inexact Block Coordinate Descent}

As mentioned previously, the normalizing terms $\| \bb_i \|$ in the objective function are troublesome.
We deal with them by considering the change of variable
\[
\bB = \bD \bC, ~~
\bC = \begin{bmatrix}
\bc_1^\top \\ \vdots \\ \bc_N^\top
\end{bmatrix}, ~~
\bD = \begin{bmatrix}
d_1 \\
& \ddots \\
& & d_N
\end{bmatrix}, ~~
d_i > 0, ~~
\bc_i \in \setU := \{ \bc \in \Rbb^N \mid  \| \bc \| = 1 \}, ~ \forall i.
\]
Applying the above transformation to Formulation 3 leads to the following reformulation
\beq
\begin{aligned} \label{eq:pr_sisal_1}
	\min_{\bC \in \Rbb^{N \times N}, \bd \in \Rbb^N } & ~ - \log | \det(\bC) | - \sum_{i=1}^N \log d_i - \frac{1}{T} \sum_{t=1}^T \sum_{i=1}^N \log \Phi( \bc_i^\top \bar{\by}_t ) \\
	{\rm s.t.} & ~ \bC^\top \bd = \bp, ~ \bC \in \setU^N,
\end{aligned}
\eeq
where, for convenience, we denote $\bar{\by}_t  = \by_t /\sigma$, $\setU^N= \{ \bC = [~ \bc_1, \ldots, \bc_N ~]^\top \mid \bc_i \in \setU ~ \forall i \}$, and $ \bd = (d_1,\ldots,d_N) $;
note ${\rm dom}\, (-\log) = \Rbb_{++}$.
The upshot of the reformulation in \eqref{eq:pr_sisal_1} is that the normalizing terms disappear.
The new challenges are that we are now faced with unit modulus constraints, and  handling both the equality constraint $\bC^\top \bd = \bp$   and the unit modulus constraints is difficult.
We make a compromise by considering a penalized alternation of problem \eqref{eq:pr_sisal_1}
\beq
\begin{aligned} \label{eq:pr_sisal}
	\min_{\bC \in \setU^N, \bd \in \Rbb^N } & ~ F_\eta(\bC,\bd): = - \log | \det(\bC) | - \sum_{i=1}^N \log d_i - \frac{1}{T} \sum_{t=1}^T \sum_{i=1}^N \log \Phi( \bc_i^\top \bar{\by}_t ) + \eta \| \bC^\top \bd - \bp \|^2
\end{aligned}
\eeq
for a given penalty parameter $\eta > 0$ that is presumably large.
Observe that $F_\eta$ is convex in $\bd$, and non-convex in $\bC$.

We employ a block coordinate descent (BCD) strategy to handle problem \eqref{eq:pr_sisal}.
The first layer of our algorithm is shown in Algorithm~\ref{alg:prsisal}.
We minimize $F_\eta$ over $\bC$ and $\bd$ in an alternating fashion.
To be more precise, the minimization $F_\eta$ over $\bC \in \setU^N$ is only approximate since the problem is non-convex.
Moreover, we gradually increase $\eta$.
By experience, graduating increasing $\eta$ is better than applying a large fixed $\eta$.
The second layer of our design deals with the computations of the coordinate minimizers in Steps 5--6 of Algorithm~\ref{alg:prsisal}, which is detailed next.

\begin{algorithm}[hbt]
	\caption{Pr-SISAL, an inexact BCD algorithm for the altered problem \eqref{eq:pr_sisal} of Formulation 3} \label{alg:prsisal}
	\begin{algorithmic}[1]
		\STATE \textbf{given:} an invertible starting point $\bB^0$, a starting penalty value $\eta > 0$, $c > 1$, and a rule for increasing $\eta$
		\STATE $k= 0$, $\bd^0 = ( \| \bb_1^0 \|, \ldots, \| \bb_N^0 \| )$, $\bC^0 = [~ \bb_1^0/d_1^0, \ldots, \bb_N^0/d_N^0 ~]^\top$
		\STATE {\bf repeat}
		\STATE \hspace{1em} {\bf repeat}
		\STATE \hspace{2em} $\bd^{k+1} = \arg \min_{\bd \in \Rbb^N} F_\eta(\bC^k,\bd)$ by Algorithm~\ref{alg:prsisal_d} with $\bd^k$ as the starting point
		\STATE \hspace{2em} $\bC^{k+1} \approx \arg \min_{\bC \in \setU^N} F_\eta(\bC,\bd^{k+1})$ by Algorithm~\ref{alg:prsisal_C} with $\bC^k$ as the starting point
		\STATE \hspace{2em} $k= k+1$
		\STATE \hspace{1em} {\bf until} a stopping rule is satisfied
		\STATE \hspace{1em} $\eta = \eta \, c$			
		\STATE {\bf until} a stopping rule is satisfied
		\STATE {\bf output:} $\bB^k = \bD^k \bC^k$, where $\bD^k = {\rm Diag}(\bd_k)$
	\end{algorithmic}
\end{algorithm}


\subsection{Coordinate Minimization Over $\bd$}

Let us first consider the coordinate minimization over $\bd$ in Step 5 of Algorithm~\ref{alg:prsisal}.
The problem amounts to solving
\beq \label{eq:min_d}
\min_{\bd \in \Rbb^N} ~ f(\bd): = \underbrace{\eta \| \bC^\top \bd - \bp \|^2}_{:=f_0(\bd)}  \underbrace{ \textstyle-  \sum_{i=1}^N \log(d_i)}_{:= f_1(\bd)}.
\eeq
The above problem is convex.
It also falls into the scope of proximal gradient methods (cf. Section~\ref{sect:lsb_pg}), with Lipschitz continuous $\nabla f_0$.
We employ the (standard) extrapolated proximal gradient method to compute the solution to problem \eqref{eq:min_d}.
The algorithm is shown in Algorithm~\ref{alg:prsisal_d}.
Note that
\beq \label{eq:prox_log}
{\rm prox}_{\mu^{-1} f_1}(\bd) = \left( \tfrac{d_1 + \sqrt{d_1^2 + 4/\mu}}{2}, \cdots,  \tfrac{d_N + \sqrt{d_N^2 + 4/\mu}}{2}    \right).
\eeq

{\begin{algorithm}[hbt]
		\caption{an extrapolated proximal gradient algorithm for $\min_{\bd \in \Rbb^N} F_\eta(\bC,\bd)$} \label{alg:prsisal_d}
		\begin{algorithmic}[1]
			\STATE \textbf{given:} a starting point $\bd^0$;  and an extrapolation sequence $\{ \alpha_k \}$, typically the FISTA sequence \cite{beck2017first}
			\STATE $k= 0$, $\bd^{-1} = \bd^0$,
			\STATE $\mu = 2 \eta \sigma_{\rm max}(\bC)^2$, where $\sigma_{\rm max}(\bC)$ is the largest singular value of $\bC$
			\STATE {\bf repeat}
			\STATE \hspace{1em} $\bd_{\rm ex}^k = \bd^k + \alpha_k ( \bd^k - \bd^{k-1} )$
			\STATE \hspace{1em} $\bd^{k+1} = {\rm prox}_{\mu^{-1} f_1}( \bd^k_{\rm ex} - \mu^{-1} \nabla f_0(\bd^k_{\rm ex}) )$; $f_0$ is given in \eqref{eq:min_d}; ${\rm prox}_{\mu^{-1} f_1}$ is given in \eqref{eq:prox_log}
			\STATE \hspace{1em} $k= k+1$
			\STATE {\bf until} a stopping rule is satisfied
			\STATE {\bf output:} $\bd^k$
		\end{algorithmic}
\end{algorithm}}

\subsection{Coordinate Minimization Over $\bC$}

Next, consider the coordinate minimization over $\bC$. The problem can be presented as
\beq
\min_{\bC \in \Rbb^{N \times N}} ~ f(\bC):= \underbrace{ - \log | \det(\bC) |  - \frac{1}{T} \sum_{t=1}^T \sum_{i=1}^N \log \Phi( \bc_i^\top \bar{\by}_t ) + \eta \| \bC^\top \bd - \bp \|^2}_{:=f_0(\bC)} + \underbrace{ \mathbb{I}_{\setU^N}(\bC) }_{:= f_1(\bC)}
\eeq
We begin by considering the proximal gradient method:
\beq
\bC^{k+1} = {\rm prox}_{\mu_k^{-1} f_1}( \bC^k - \mu_k^{-1} \nabla f_0(\bC^k) ) = \Pi_{\setU^N}( \bC^k - \mu_k^{-1} \nabla f_0(\bC^k) ),
\eeq
where $\mu_k > 0$ is chosen such that the sufficient decrease condition is satisfied, and it is done by the backtracking line search (cf. \eqref{eq:backtrack}); we have
\[
\Pi_{\setU^N}(\bC) = [~ \Pi_{\setU}(\bc_1), \ldots, \Pi_{\setU}(\bc_N) ~]^\top, \quad
\Pi_{\setU}(\bc) = \left\{
\begin{array}{ll}
\bc/ \| \bc \| & \text{if $\bc \neq \bzero$} \\ \text{any $\bu \in \setU$} & \text{if $\bc = \bzero$}
\end{array}
\right.
\]
The method, by operations, is the same as the standard proximal gradient method.
But the problem does not fall within the scope of the stationarity-guaranteed LSB-PG framework, because $\setU^N$ is non-convex.
We adopt this method mostly based on practicality:
It is simple, and
the same method or similar methods have been used in practice \cite{boumal2016nonconvex,tranter2017fast,shao2018minimum}, with reasonable results demonstrated.
Moreover, as a supporting argument, the method is shown to be equipped with some stationarity guarantee under the assumption of Lipschitz continuous $\nabla f_0$ \cite{tranter2017fast}.

The above method is just a vanilla version of our actual algorithm.
There is a practical issue: the computation of $\nabla f_0$ is expensive, and the direct use of the proximal gradient method can be slow in terms of the runtimes.
To give an idea, let us show $\nabla f_0$:
\[
\nabla f_0(\bC) = - \bC^{-\top} - \frac{1}{T} \sum_{t=1}^T
\begin{bmatrix}
\frac{1}{\Phi(\bc_1^\top \bar{\by}_t)} \frac{1}{\sqrt{2\pi}} e^{- (\bc_1^\top \bar{\by}_t)^2/2} \bar{\by}_t^\top \\
\vdots \\
\frac{1}{\Phi(\bc_N^\top \bar{\by}_t)} \frac{1}{\sqrt{2\pi}} e^{- (\bc_N^\top \bar{\by}_t)^2/2} \bar{\by}_t^\top
\end{bmatrix}
+ 2 \eta \, \bd (\bC^\top \bd - \bp )^\top.
\]
We see that computing $\nabla f_0$ requires evaluating $\Phi$ for a number of $NT$ times (recall that $T$ is large in practice).
The function  $\Phi$ does not have a closed form and is evaluated by a numerical method.
While this should not be an issue when we are required to call $\Phi$ a few times, the problem here requires us to evaluate $\Phi$ numerous times (and at every iteration).

To reduce the number of times $\Phi$ is called, and thereby alleviate the computational burden,
we consider a combination of the majorization-minimization (MM) and proximal gradient method.
Recall the idea of MM: i) build a surrogate of $f$ by finding a majorant  $g(\bC,\tilde{\bC})$ of $f$ at $\tilde{\bC}$, i.e.,
$f(\bC) \leq g(\bC,\tilde{\bC})$ for all $\bC,\tilde{\bC}$, and $f(\bC) = g(\bC,\bC)$;
ii) handle the problem by recursively solving $\bC^{k+1} = \min_{\bC} g(\bC,\bC^k)$.
Consider the following fact.

\begin{Fact}[\cite{shao2021divide} and the references therein] \label{fact:mm}
	It holds that, for any $\tilde{x} \in \Rbb$,
	\[
	-\log \Phi(x) \leq g(x, \tilde{x}) := \frac{1}{2} | x + w(\tilde{x}) |^2 + r(\tilde{x}),
	\]
	where $r(\tilde{x})$ does not depend on $x$;
	\[
	w(\tilde{x}) = - \tilde{x} - \frac{1}{\Phi(\tilde{x})} \frac{1}{\sqrt{2\pi}} e^{- \tilde{x}^2 / 2 }.
	\]
	Also, we have $g(x,x) = -\log \Phi(x)$.
\end{Fact}

Let us apply Fact~\ref{fact:mm} to build a majorant of $f_0$:
\beq \label{eq:g0}
g_0(\bC,\tilde{\bC}) =  - \log | \det(\bC) |  + \frac{1}{2T} \sum_{t=1}^T \sum_{i=1}^N
\left|
\bc_i^\top \bar{\by}_t - w( \tilde{\bc}_i^\top \bar{\by}_t )
\right|^2
+ \eta \| \bC^\top \bd - \bp \|^2 + r(\tilde{\bC}),
\eeq
for some $r$ that does not depend on $\bC$.
Also, let $g(\bC,\tilde{\bC})= g_0 (\bC,\tilde{\bC}) + f_1(\bC)$, which is a majorant of $f$.
We carry out MM, in an inexact sense, by approximating  $\bC^{k+1} = \arg \min_{\bC} g(\bC,\bC^k)$ via the proximal gradient method.
By doing so, we hope that the number of times $\Phi$ is called can be reduced:
the evaluations of $\Phi$ happen in the majorant construction step \eqref{eq:g0}, but not in the (more intensively operating) proximal gradient iterations.
Our high-level algorithm description is complete, and the algorithm is shown below.
Note that the actual proximal gradient method we employ is extrapolated.

\begin{algorithm}[hbt]
	\caption{a combined MM and extrapolated proximal gradient algorithm for $\min_{\bC \in \setU^N} F_\eta(\bC,\bd)$} \label{alg:prsisal_C}
	\begin{algorithmic}[1]
		\STATE \textbf{given:} an invertible starting point $\bC^0$;  and an extrapolation sequence $\{ \alpha_k \}$, typically the FISTA sequence \cite{beck2017first}
		\STATE $k= 0$,
		\STATE {\bf repeat} \hspace{1em} \% MM iterations
		\STATE \hspace{1em} compute $w( (\bc_i^k)^\top \bar{\by}_t )$ for all $i,t$
		\STATE \hspace{1em} $l= 0$, $\bC^{k,-1} = \bC^{k,0} = \bC^k$
		\STATE \hspace{1em} {\bf repeat} \hspace{1em} \% extrapolated proximal gradient iterations
		
		\STATE \hspace{2em} $\bC_{\rm ex}^{k,l} = \bC^{k,l} + \alpha_l ( \bC^{k,l} - \bC^{k,l-1} )$
		\STATE \hspace{2em} $\bC^{k,l+1} = \Pi_{\setU^N}( \bC_{\rm ex}^{k,l} - \mu_{k,l}^{-1} \nabla g_0( \bC_{\rm ex}^{k,l}, \bC^k ))$, where $\mu_{k,l}$ is chosen such that
		\[ g_0(\bC^{k,l+1},\bC^k ) \leq g_0(\bC_{\rm ex}^{k,l},\bC^k ) + \langle \nabla g_0(\bC_{\rm ex}^{k,l},\bC^k ), \bC^{k,l+1} - \bC_{\rm ex}^{k,l} \rangle + \tfrac{\mu_{k,l}}{2} \| \bC^{k,l+1} - \bC_{\rm ex}^{k,l} \|^2 \]
		\hspace{2em} (i.e., sufficient decrease)	is satisfied, and it is done by the backtracking line search; \\
		\hspace{2em}  $g_0$  is given in \eqref{eq:g0}
		\STATE \hspace{2em} $l= l+1$

		\STATE \hspace{1em} {\bf until} a stopping rule is satisfied
		\STATE \hspace{1em} $\bC^{k+1} = \bC^{k,l}$
		\STATE \hspace{1em} $k= k+1$
		\STATE {\bf until} a stopping rule is satisfied
		\STATE {\bf output:} $\bC^k$
	\end{algorithmic}
\end{algorithm}

\section{Numerical Results}

Now we proceed to numerical results.
While we focused on giving a novel explanation of SISAL,
the study itself showed new possibilities which we would like to examine by numerical experiments.
The most interesting one is the approximate ML estimator in Formulation 3, which resembles a SISAL variant that adopts a probabilistic penalty term.
This probabilistic SISAL does not have the regularization parameter $\lambda$,
and we want to see how well it works compared to SISAL (which requires tuning $\lambda$).
Also we are interested in the hinge-square SISAL variant in Formulation 4, in terms of runtimes.

\subsection{Settings of the Algorithms}
\label{sect:sim_settings}

The implementations of the hinge-square and probabilistic SISAL formulations in Formulations 4 and 3 are accomplished by Algorithms \ref{alg:h2sisal} and \ref{alg:prsisal}, respectively.
For convenience, Algorithms \ref{alg:h2sisal} and \ref{alg:prsisal} will be called H$^2$-SISAL and Pr-SISAL, respectively, in the sequel.
We first specify the dimensionality reduction (DR) preprocessing, which is required by the SISAL algorithms.
The standard PCA is used to perform DR.
To be specific, let $\by_1,\ldots,\by_T \in \Rbb^M$ be the data points.
We compute $\hat{\bR}_{yy} = \frac{1}{T} \sum_{t=1}^T \by_t \by_t^\top$, compute the $N$-principal eigenvector matrix $\bU \in \Rbb^{M \times N}$ of $\hat{\bR}_{yy}$, and take $\tilde{\by}_t = \bU^\top \by_t \in \Rbb^N$ as the dimension-reduced data points.
Pr-SISAL or H$^2$-SISAL is then applied to $\tilde{\by}_1,\ldots,\tilde{\by}_T$ to get an estimate of $\tilde{\bA}_0=  \bU^\top \bA_0$, and we use the relation $\bA_0 = \bU \tilde{\bA}_0$ to form the estimate of $\bA_0$.
In this connection, it is worth noting that, for the case of $M \geq N+1$, we can also estimate the noise power $\sigma^2$ from $\hat{\bR}_{yy}$, specifically, by taking the $(N+1)$th eigenvalue of $\hat{\bR}_{yy}$ as the estimate of $\sigma^2$; this is a commonly-used trick in statistical signal processing \cite[Chapter~4.5]{stoica2005spectral}.

The settings of Pr-SISAL in Algorithm \ref{alg:prsisal} are as follows.
The vector $\bp$ is estimated by \eqref{eq:p_est}.
The starting point is generated by expanded vertex component analysis (VCA), a built-in function of SISAL and a slight modification of the output by the VCA algorithm \cite{Nascimento2005}.
We set the initial value of $\eta$ to $1$ and set $c= 5$.
We stop the inner loop (Steps 4--8) if ${\rm rc}(\bB^{k+1},\bB^k):= \| \bB^{k+1} - \bB^{k} \|/\| \bB^k \| \leq  10^{-7}$ ( rc stands for relative change) or if the number of inner loops  exceeds  $4 \times 10^{5}$.
We stop the outer loop if the number of outer loops exceeds $10$.
For the sub-algorithm Algorithm \ref{alg:prsisal_d}, we stop  if ${\rm rc}(\bd^{k+1},\bd^k) \leq 10^{-5}$.
For the sub-algorithm Algorithm \ref{alg:prsisal_C}, we stop the MM loop and the proximal gradient loop if ${\rm rc}(\bC^{k+1},\bC^k) \leq 10^{-5}$ and ${\rm rc}(\bC^{k,l+1},\bC^{k,l}) \leq 10^{-3}$, respectively.
The extrapolation sequence $\{ \alpha_k \}$ in Algorithms \ref{alg:prsisal_d} and \ref{alg:prsisal_C} is chosen as the (standard) FISTA sequence \cite{beck2017first}.

The settings of H$^2$-SISAL in Algorithm \ref{alg:h2sisal} are as follows.
We choose $\bp= (\bY^\top)^\dag \bone$.
The starting point is generated by expanded VCA.
The FISTA extrapolation sequence is used.
We stop Algorithm \ref{alg:h2sisal} if ${\rm rc}(\bB^{k+1},\bB^k) \leq 10^{-6}$.

We will benchmark Pr-SISAL and H$^2$-SISAL against SISAL itself, VCA~\cite{Nascimento2005}, ISA-PRISM and VIA-PRISM~\cite{PRISM2021}.
SISAL and VCA have open source codes, and we use them directly.
The stopping rule of SISAL is that the number of iterations exceeds $250$.
ISA-PRISM is an importance sampling scheme for implementing the ML estimator \eqref{eq:mle}, and VIA-PRISM is a variational inference approximation scheme for the ML estimator \eqref{eq:mle}.
We run ISA-PRISM only for small $N$, due to its demanding computational cost to achieve reasonable performance for large $N$.
We stop ISA-PRISM when the number of iterations exceeds $100$, and we use rejection sampling, with $500$ initial samples, to implement ISA-PRISM.
We stop VIA-PRISM when the number of iterations exceeds 500.
Also, our VIA-PRISM implementation has some differences from that in the original work \cite{PRISM2021};
we replace the optimization algorithm for the variational variables, Algorithm 1 in \cite{PRISM2021}, with a projected gradient algorithm, which was found to be more efficient.

\subsection{Comparisons of SISAL, H$^2$-SISAL and Pr-SISAL By Simulations}
\label{sect:syth_sim}

We conduct our simulations by the following way.
We generate the data points $\by_1,\ldots,\by_T$ by the model in \eqref{eq:prob_model}, i.e., $\by_t = \bA_0 \bs_t + \bv_t$, where the $\bs_t$'s are i.i.d. uniform distributed on the unit simplex; the $\bv_t$'s are i.i.d. Gaussian with mean zero and covariance $\sigma^2 \bI$.
In addition, for each simulation trial, $\bA_0$ is drawn from an element-wise i.i.d. $[0,1]$ distribution;
we also restrict the condition number of the admitted $\bA_0$ to be no greater than $100$.
We use a number of $100$ simulation trials to evaluate the mean square error (MSE)
\[
{\sf MSE}(\bA_0,\hat{\bA}) = \min_{\bm P \in \mathcal{P} } \frac{1}{MN} \| \bA_0 - \hat{\bA} \bP \|^2,
\]
where $\hat{\bA}$ denotes an estimate of $\bA_0$ by some algorithm; $\mathcal{P}$ is the set of all permutation matrices on $\Rbb^{N \times N}$.
We should also note that the signal-to-noise ratio (SNR) is defined as
\[
{\sf SNR} = \frac{\frac{1}{T} \sum_{t=1}^T \| \bA_0 \bs_t \|^2 }{ M \sigma^2 }
\]

\begin{figure}[p]
	\centering
	\begin{minipage}[b]{.75\textwidth}
		\centering
		\includegraphics[width=\linewidth]{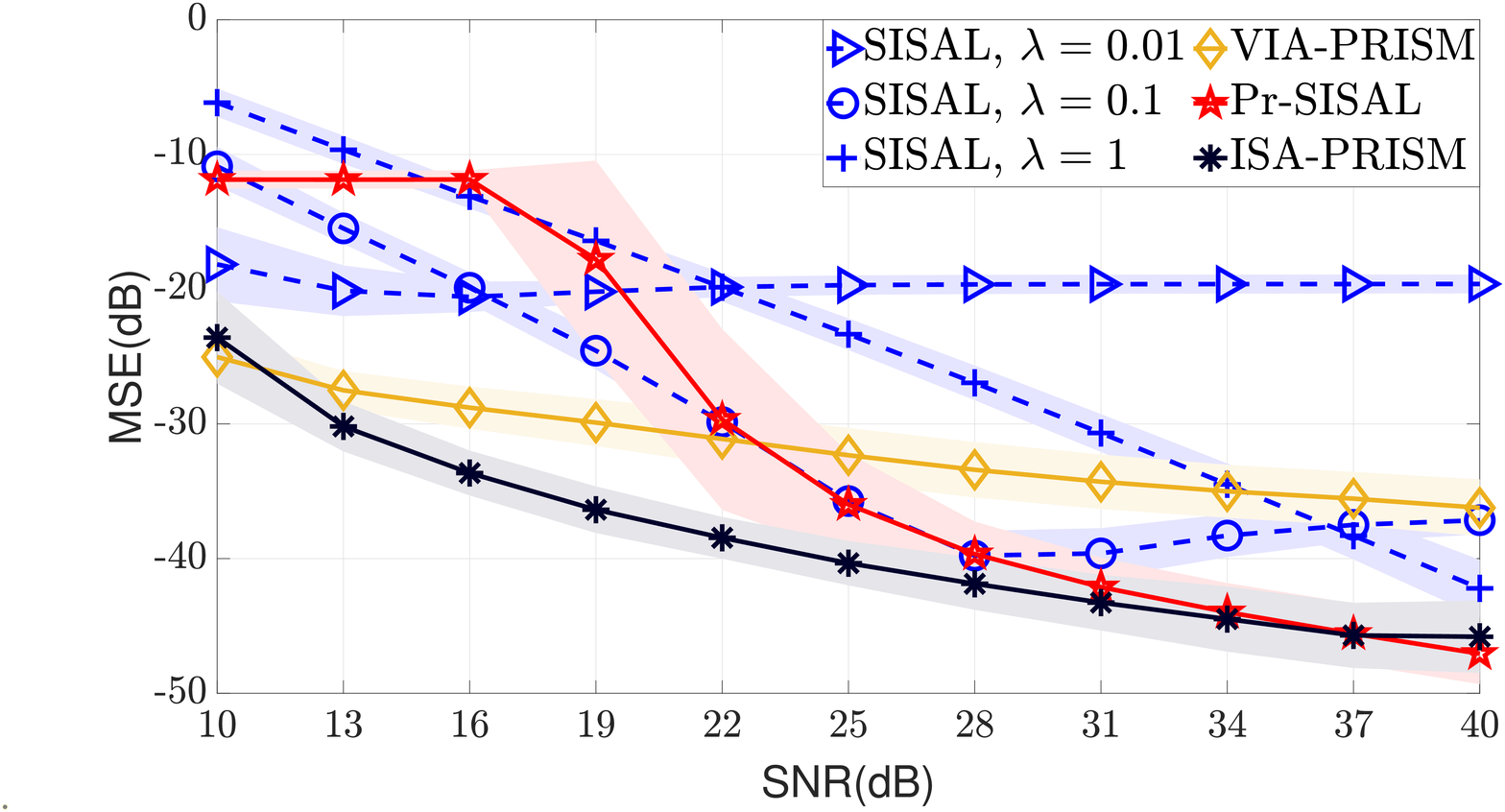}
		(a) $M = 10$, $N = 5$
	\end{minipage}
	\begin{minipage}[b]{.75\textwidth}
		\centering
		\includegraphics[width=\linewidth]{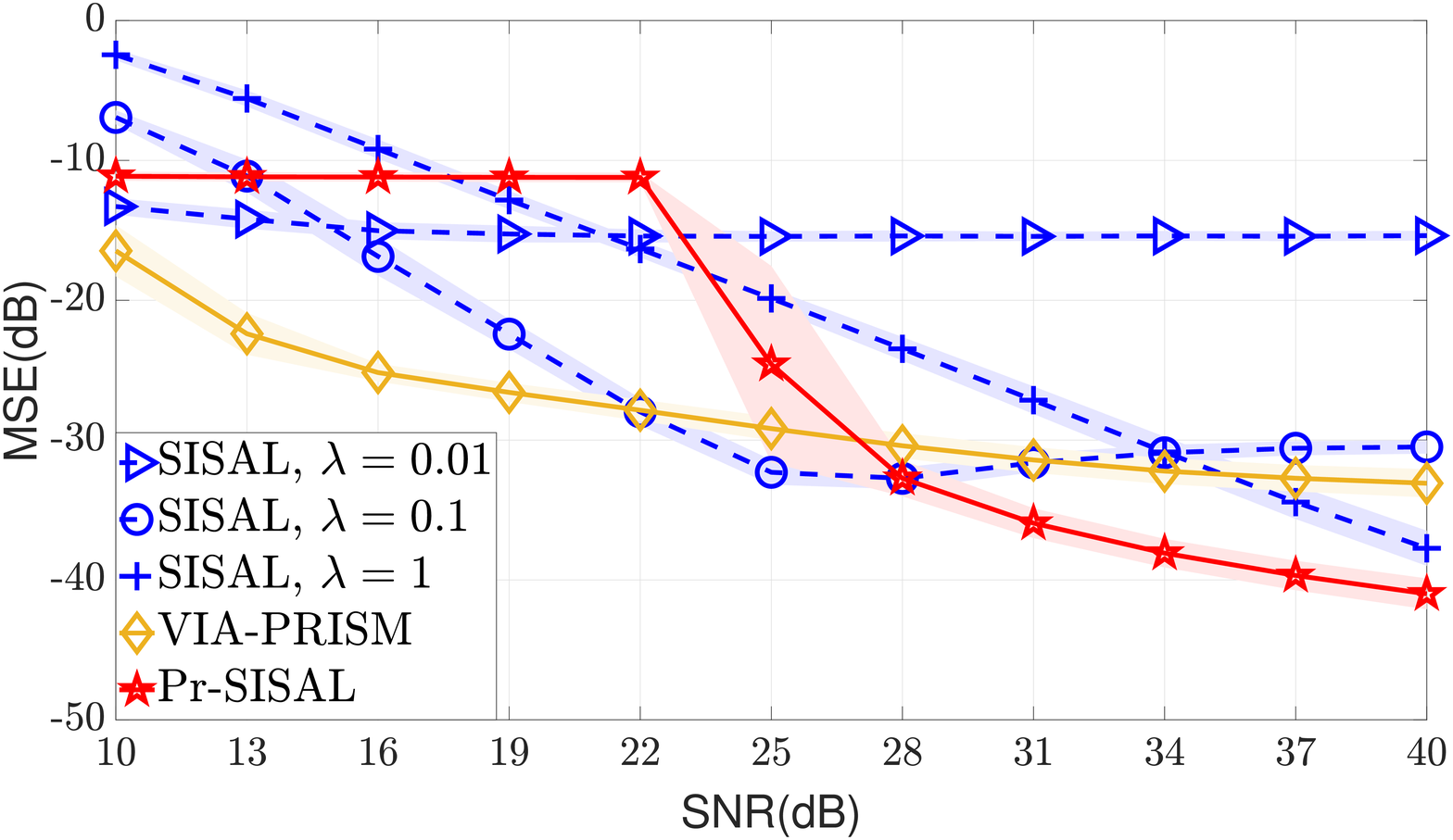}
		(b) $M = 20$, $N = 10$
	\end{minipage}
	\begin{minipage}[b]{.75\textwidth}
		\centering
		\includegraphics[width=\linewidth]{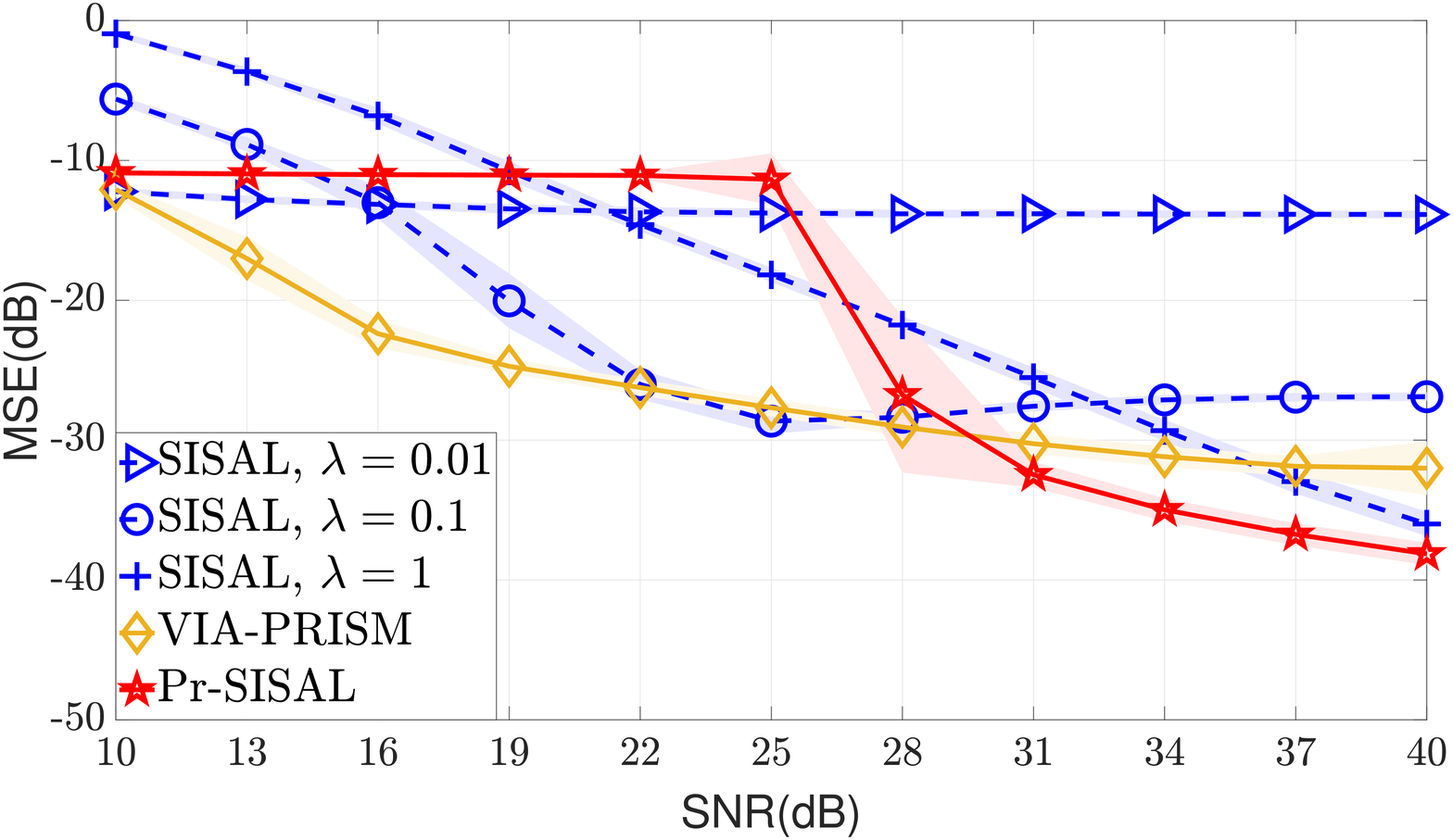}
		(c) $M = 30$, $N = 15$
	\end{minipage}
	\caption{Comparison of Pr-SISAL, SISAL and VIA-PRISM. The lines are the average MSEs, while the shaded areas show the standard deviations of the MSEs.}
	\label{fig:PrSISAL}	
\end{figure}

Fig.~\ref{fig:PrSISAL} compares Pr-SISAL and SISAL for various values of $(M,N)$ and for $T=1,000$.
Our observations are as follows.
First, the recovery performance behaviors of SISAL vary from one choice of $\lambda$ to another.
There is no single $\lambda$ that works best for all SNRs, which suggests the need for parameter tuning in practice.
Second, Pr-SISAL performs unsatisfactorily for low SNRs, particularly when compared to VIA-PRISM. But we also see that the performance of Pr-SISAL improves drastically as the SNRs are greater than certain thresholds.
Also, for $(M,N)= (10,5)$, Pr-SISAL achieves performance close to the ML estimator by ISA-PRISM when the SNR is high enough.
These results indicate that Pr-SISAL is a good estimator for the high SNR regime.

\begin{figure}[p]
	\centering
	\begin{minipage}[b]{.75\textwidth}
		\centering
		\includegraphics[width=\linewidth]{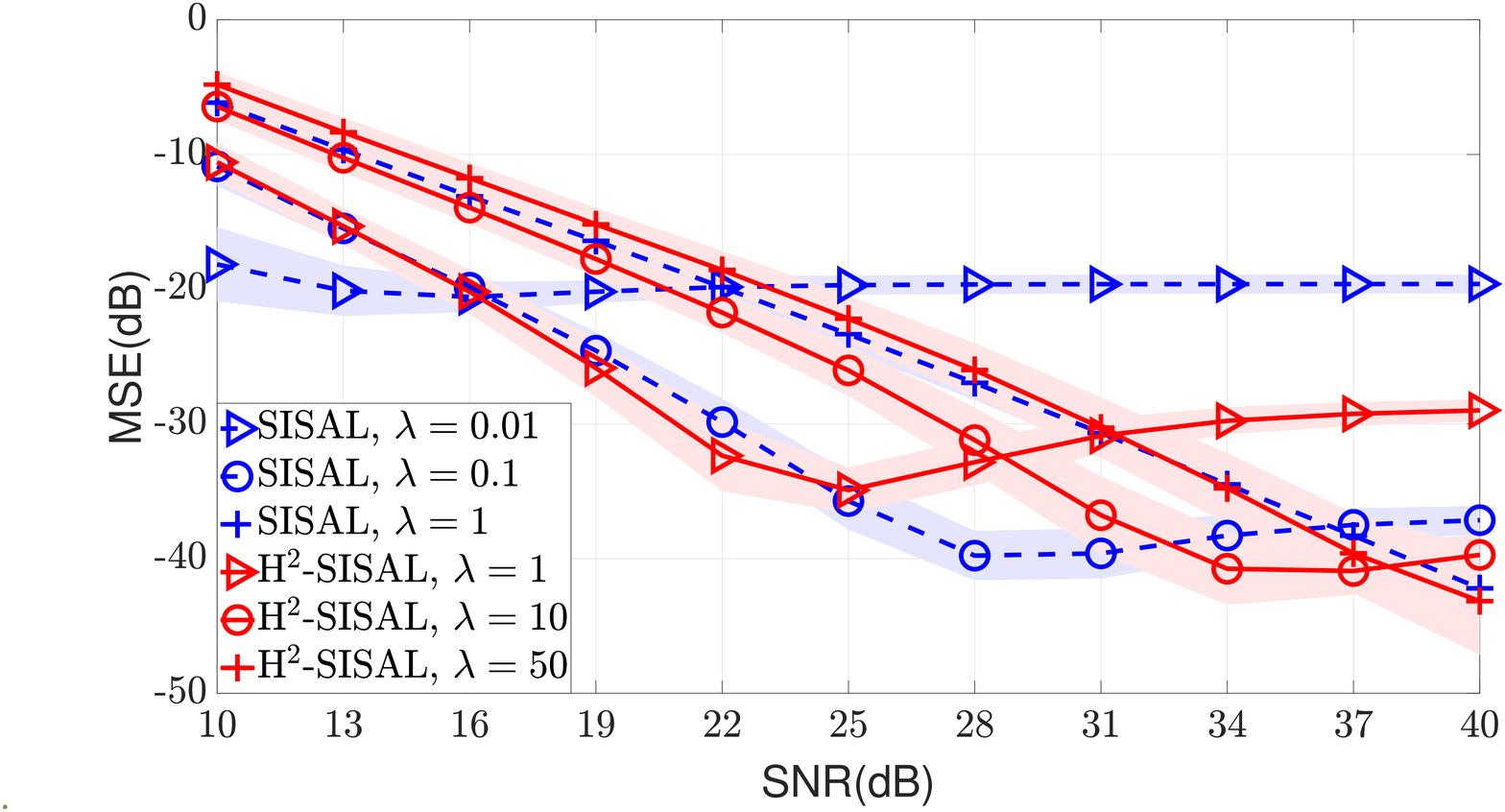}
		(a) $M = 10$, $N = 5$
	\end{minipage}
	\begin{minipage}[b]{.75\textwidth}
		\centering
		\includegraphics[width=\linewidth]{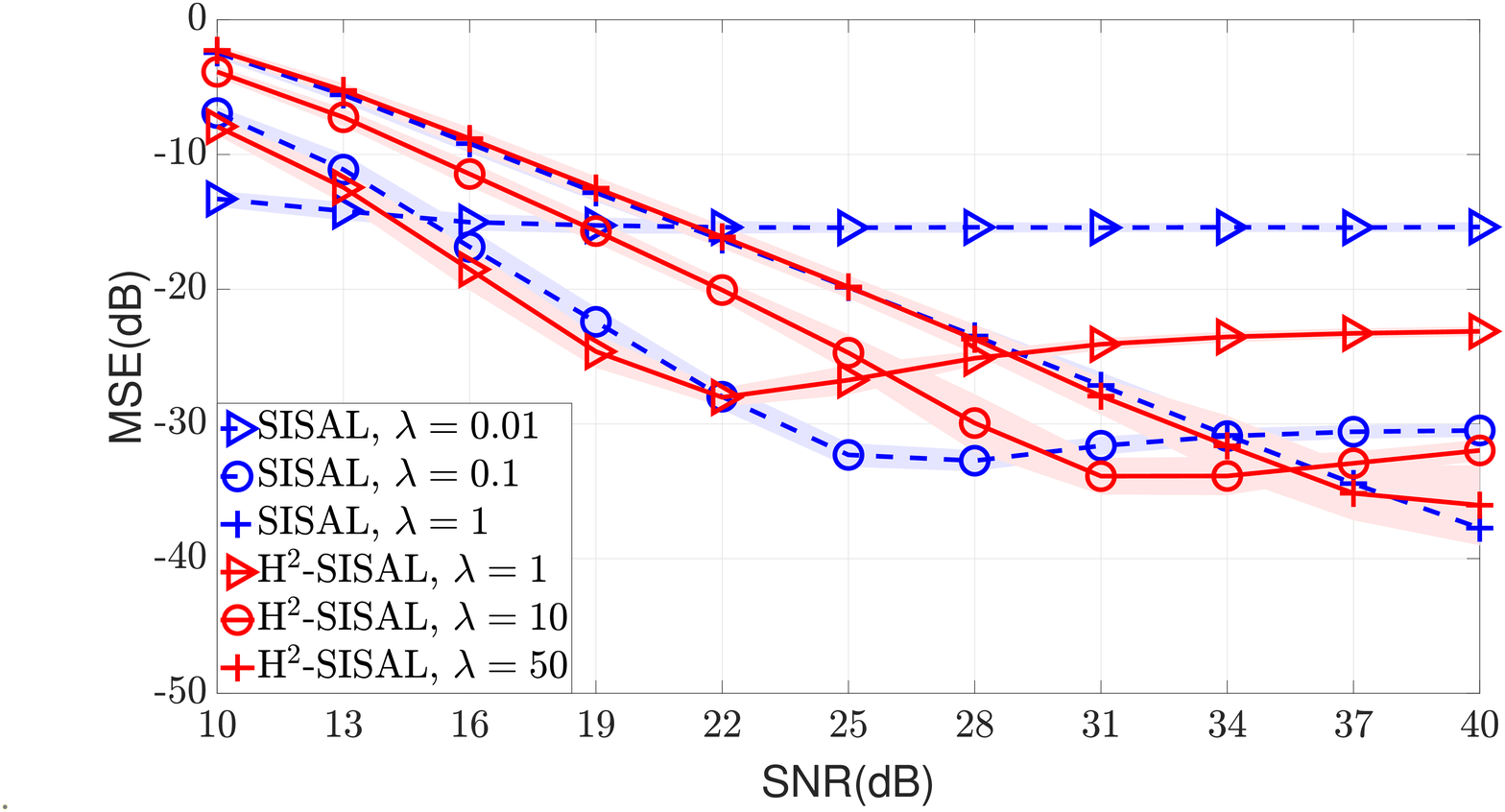}
		(b) $M = 20$, $N = 10$
	\end{minipage}
	\begin{minipage}[b]{.75\textwidth}
		\centering
		\includegraphics[width=\linewidth]{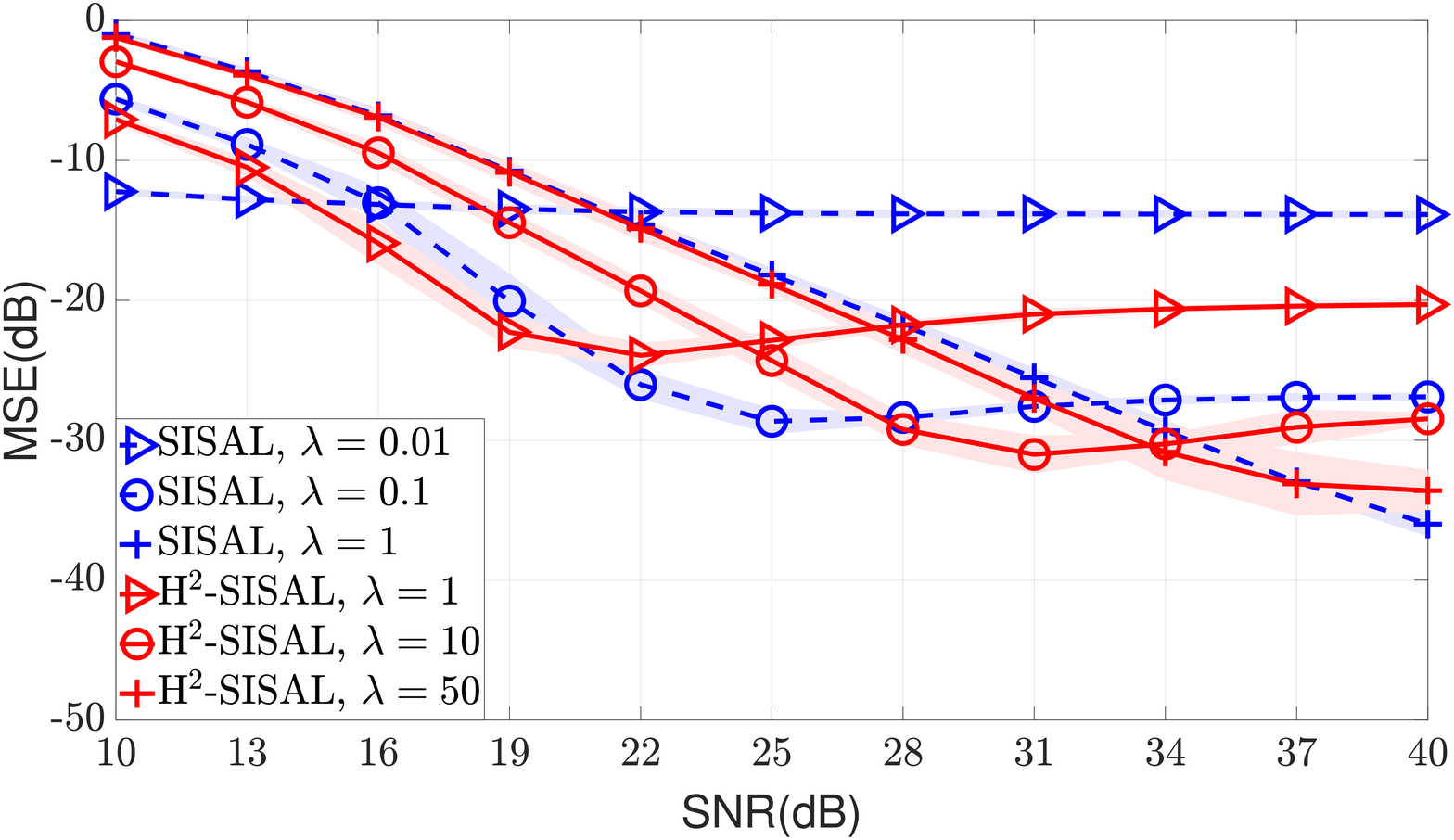}
		(c) $M = 30$, $N = 15$
	\end{minipage}
	\caption{Comparison of H$^2$-SISAL and SISAL.}
	\label{fig:H2SISAL}	
\end{figure}

Fig.~\ref{fig:H2SISAL} compares H$^2$-SISAL and SISAL under the same settings as above.
We see that H$^2$-SISAL works reasonably and is comparable to SISAL.
Also, H$^2$-SISAL behaves differently for different regularization parameters $\lambda$, which suggests that H$^2$-SISAL requires parameter tuning in practice (just like SISAL).

We move on to the comparison of computational efficiency.
Tables~\ref{Tab:runtimeforSISAL+H2SISAL}--\ref{Tab:runtimeUnderdifferentT} illustrate some runtime results.
The runtimes were measured on a small server with the Intel  Core i7-5820K CPU processor and 64GB memory, and with implementations using MATLAB 2019a.
${\rm H}^2$-SISAL is seen to run faster than SISAL.
Pr-SISAL, in comparison, is slow, although this is so far the best algorithm we can build for the difficult formulation of probabilistic SISAL.
The reader will see in the extra simulation results in Appendix~\ref{sect:add_sim} that the proximal gradient method for tackling SISAL and H$^2$-SISAL is even slower for  probabilistic SISAL.


\begin{table}[!hbt]
	\centering
	\caption{Average runtimes (in sec.) of SISAL, H$^2$-SISAL, Pr-SISAL and VIA-PRISM. $T=1,000$, ${\sf SNR}= 30$dB.}
	\small
	\renewcommand{\arraystretch}{1.2}
	\begin{tabular}{c||c|c|c}
		\hline\hline
		$(M,N)$	&$(10, 5)$ & $(20,10)$ & $(30,15)$  \\
		\hline \hline
		SISAL, ${\lambda = 0.1} $  			  & 0.078 & 0.129 & 0.162    \\
		\hline
		${\rm H}^2$-SISAL, ${\lambda =10.0} $ & 0.033 & 0.066 & 0.132    \\
		\hline
		Pr-SISAL  & 8.336 & 21.854 & 42.785    \\
		\hline
		VIA-PRISM & 0.632 & 0.974 & 1.212 \\
		\hline \hline
	\end{tabular}
	\label{Tab:runtimeforSISAL+H2SISAL}
\end{table}


\begin{table}[!hbt]
	\centering
	\caption{Average runtimes (in sec.) of SISAL, H$^2$-SISAL, Pr-SISAL and VIA-PRISM. $(M,N)= (20,10)$, ${\sf SNR}= 30$dB.}
	\small
	\renewcommand{\arraystretch}{1.2}
	\begin{tabular}{c||c|c|c|c|c|c|c|c}
		\hline\hline
		$T$  &  1000  &  2000  &  3000  &  4000  &  5000  &  6000  &  7000  &  8000   \\
		\hline \hline
		SISAL, ${\lambda = 0.1} $  		& 0.119  & 0.201  & 0.295  & 0.353 & 0.401 & 0.455 & 0.539 & 0.587  \\
		\hline
		${\rm H}^2$-SISAL, ${\lambda =10.0} $ & 0.064  &  0096  &  0.139 & 0.192 & 0.230 & 0.246 & 0.281 & 0.325  \\
		\hline
		Pr-SISAL 	& 23.145  & 24.656  & 50.609 & 56.395 & 75.753 & 75.278&100.100&100.917  \\
		\hline
		VIA-PRISM  & 0.986 & 1.600 & 2.276 & 2.860 & 3.349 & 3.928 & 4.746 & 4.961 \\
		\hline \hline
	\end{tabular}
	\label{Tab:runtimeUnderdifferentT}
\end{table}

\subsection{A Semi-Real Data Experiment}

\begin{figure}[htb!]
	\centering
	~ \hfil
	\begin{minipage}[t]{.4\textwidth}
		\centering
		\includegraphics[width=.75\linewidth]{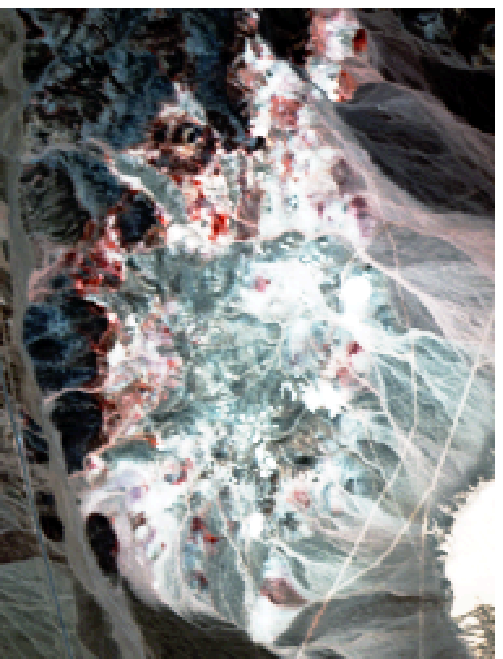} \\
		(a)  Cuprite image
	\end{minipage}
	\hfil
	\begin{minipage}[t]{.4\textwidth}
		\centering
		\includegraphics[width=.75\linewidth]{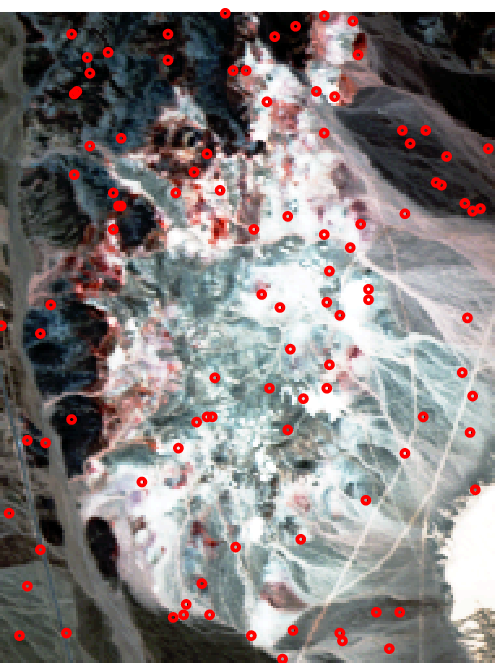} \\
		(b) Cuprite image with artificially added  outliers; red circles represent outlying pixels.
	\end{minipage}
	\hfil ~
	\caption{Cuprite image; constructed by RGB bands.}
	\label{fig:Cuprite_data}	
\end{figure}

We further test Pr-SISAL by using real data.
The application of interest is hyperspectral unmixing (HU). The real data set used to perform our experiment is the Cuprite hyperspectral image \cite{vane1993airborne}; we will simply call it Cuprite for convenience.
Cuprite is interesting in the sense that, among the popular and publicly available data sets in hyperspectral remote sensing, Cuprite is the only one that has more than $10$ materials (to our best knowledge).
Cuprite has been used to demonstrate many HU algorithms, e.g., \cite{Nascimento2005,Chan2009,MVSA,PRISM2021},
and real data experiments by Cuprite have almost become a standard.
An illustration of the Cuprite image is shown in Fig.~\ref{fig:Cuprite_data}(a). 

The settings of our experiment are as follows.
We largely follow the standard procedure in the literature \cite{Nascimento2005,MVSA,Chan2009,PRISM2021}, particularly, the one in \cite{PRISM2021}.
Some additional details are as follows.
We adopt the band selection in \cite{MVSA}.
It was argued that Cuprite is composed of $12$ materials, namely, those shown in Table \ref{Tab:CupriteExpNormal2}; we refer the reader to \cite{zhu2017hyperspectral} and the references therein for details. The ground-truth $\bA_0$ corresponds to the reference spectral responses of those materials, taken from the USGS library \cite{clark2007usgs}.
We test VCA, VIA-PRISM, SISAL, H$^2$-SISAL and Pr-SISAL.
For all the tested algorithms, we additionally do the following: we apply the data normalization preprocessing, described in Section~\ref{sect:prob_stat}, to the data points before DR;
also, for Pr-SISAL and VIA-PRISM, we estimate the noise variance $\sigma^2$ by the eigenvalue method described in Section~\ref{sect:sim_settings}.
Moreover, some of the stopping rules are modified:
We stop SISAL if the number of iterations exceeds $1,000$;
we stop the inner loop of Pr-SISAL if ${\rm rc}(\bB^{k+1},\bB^k) \leq 2 \times 10^{-7}$ or if the number of iterations exceeds $10^7$.
We evaluate the recovery performance by the spectral angle distance (SAD)
\[
{\sf SAD}(\ba_{0,i},\hat{\ba}_{\pi_i}) = \cos^{-1}\left(  \frac{ \ba_{0,i}^\top \hat{\ba}_{\pi_i} }{ \| \ba_{0,i} \|  \| \hat{\ba}_{\pi_i} \|  }   \right),
\]
where $\ba_{0,i}$ and $\hat{\ba}_i$ denote the $i$th column of $\bA_0$ and $\hat{\bA}$, respectively;
$\bm \pi =( \pi_1,\ldots,\pi_N  )$ is a set of permutation indices for $\{1,\ldots, N\}$ (i.e. $ \pi_i\in \{1,\ldots,N\} $ and $ \pi_i \neq \pi_j $ for all $ i\neq j $), obtained by
minimizing $ \sum_{i=1}^N {\sf SAD}(\ba_{0,i},\hat{\ba}_{\pi_i}) $ over all possible permutations.

Table~\ref{Tab:CupriteExpNormal2} shows the SADs of the tested algorithms.
We see that all the algorithms give reasonable SAD performance, with VCA achieving the best average SAD. We also see that SISAL and H$^2$-SISAL, with the regularization parameter tuned to $ \lambda = 0.001 $ and $ \lambda = 0.01 $, respectively, provide comparable performance to Pr-SISAL. But note that Pr-SISAL has no parameter to manually tune.
\begin{table}[!hbt]
	\centering
	\caption{SAD performances on the Cuprite dataset. The best SADs among all the tested algorithms are marked in {\bf bold}.}
	\scalebox{0.85}{
		\begin{tabular}{c||c|c|c|c|c|c|c}
			\hline \hline
			\multirow{2}{*}{\diagbox{\small{ Endmember}}{\small{Alg.}}} &  \multirow{2}{*}{VCA} & \multicolumn{2}{c|}{SISAL} & \multicolumn{2}{c|}{H$^2$-SISAL} & \multirow{2}{*}{Pr-SISAL} & \multirow{2}{*}{VIA-PRISM} \\
			\cline{3-6}
			&   & ${\lambda = 0.001} $ & ${\lambda =0.01} $ & ${\lambda =0.01} $ & ${\lambda =0.1} $ & &  \\
			\hline \hline
			Alunite		   	& 2.07 & 4.55  & 6.82  &  \textbf{1.65} & 3.83 & 3.27 & 4.54 \\
			\hline
			Andradite 	   	& 2.07 & 2.35  & 5.66  &  2.37 & 3.69 & \textbf{1.89} & 3.10  \\
			\hline
			Buddingtonite 	& \textbf{2.11} & 5.20  & 3.68  &  2.92 & 3.19 & 3.43 & 3.88 \\
			\hline
			Dumortierite 	& \textbf{2.66} & 3.25  & 8.07  &  3.32 & 6.49 & 3.51 & 3.39 \\
			\hline
			Kaolinite$_1$  	& 2.51 & 2.22  & 2.78  &  \textbf{2.16} & 3.06 & 2.67 & 3.90  \\
			\hline
			Kaolinite$_2$ 	& \textbf{1.99} & 2.48  & 7.77  &  2.29 & 6.20 & \textbf{1.99} & 2.79 \\
			\hline
			Muscovite 	   	& \textbf{2.12} & 2.80  & 3.15  &  6.07 & 4.30 & 3.64 & 2.67 \\
			\hline
			Montmorillonite & 1.74 & 2.53  & 3.88  &  1.99 & 2.77 & \textbf{1.27} & 3.22 \\
			\hline
			Nontronite 		& \textbf{1.97} & 3.81  & 2.84  &  3.03 & 3.72 & 2.75 & 3.14 \\
			\hline
			Pyrope 			& 2.10 & 1.45  & 3.93  &  1.94 & 2.76 & 1.70 & \textbf{1.32} \\
			\hline
			Sphene 			& \textbf{1.49} & 3.19  & 7.85  &  3.47 & 6.95 & 4.49 & 1.83 \\
			\hline
			Chalcedony 		& 2.86 & 3.82  & 3.85  &  3.09 & 3.38 & \textbf{1.59} & 4.35 \\
			\hline\hline
			Average SAD   	& \textbf{2.14} & 3.14  & 5.02  &  2.86 & 4.19 & 3.13 & 3.07 \\
			\hline
	\end{tabular}}
	\label{Tab:CupriteExpNormal2}
\end{table}

We also consider an experiment that puts some twist on the Cuprite data experiment. Specifically, we randomly pick some pixels and replace them with outliers; see Fig.~\ref{fig:Cuprite_data}(b) for an illustration.
Our aim is to examine how robust the algorithms are.
The experimental settings are the same as above, and additionally we randomly select $100$ pixels and replace them with randomly selected spectral responses from the USGS library \cite{clark2007usgs}.

\begin{figure}[htb!]
	\centering
	\begin{minipage}[b]{\textwidth}
		\centering
		\includegraphics[width=0.30\linewidth]{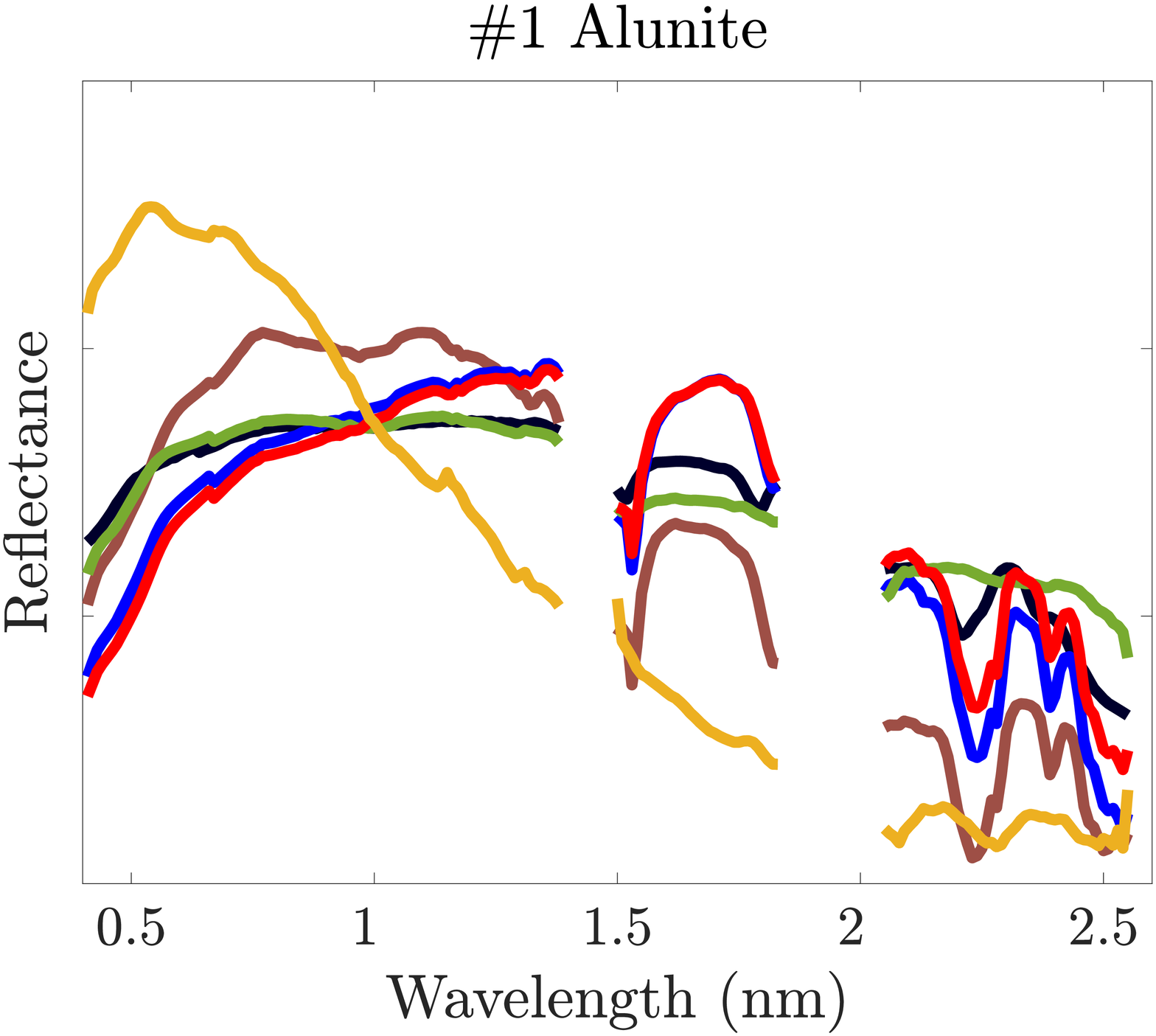}
		\includegraphics[width=0.30\linewidth]{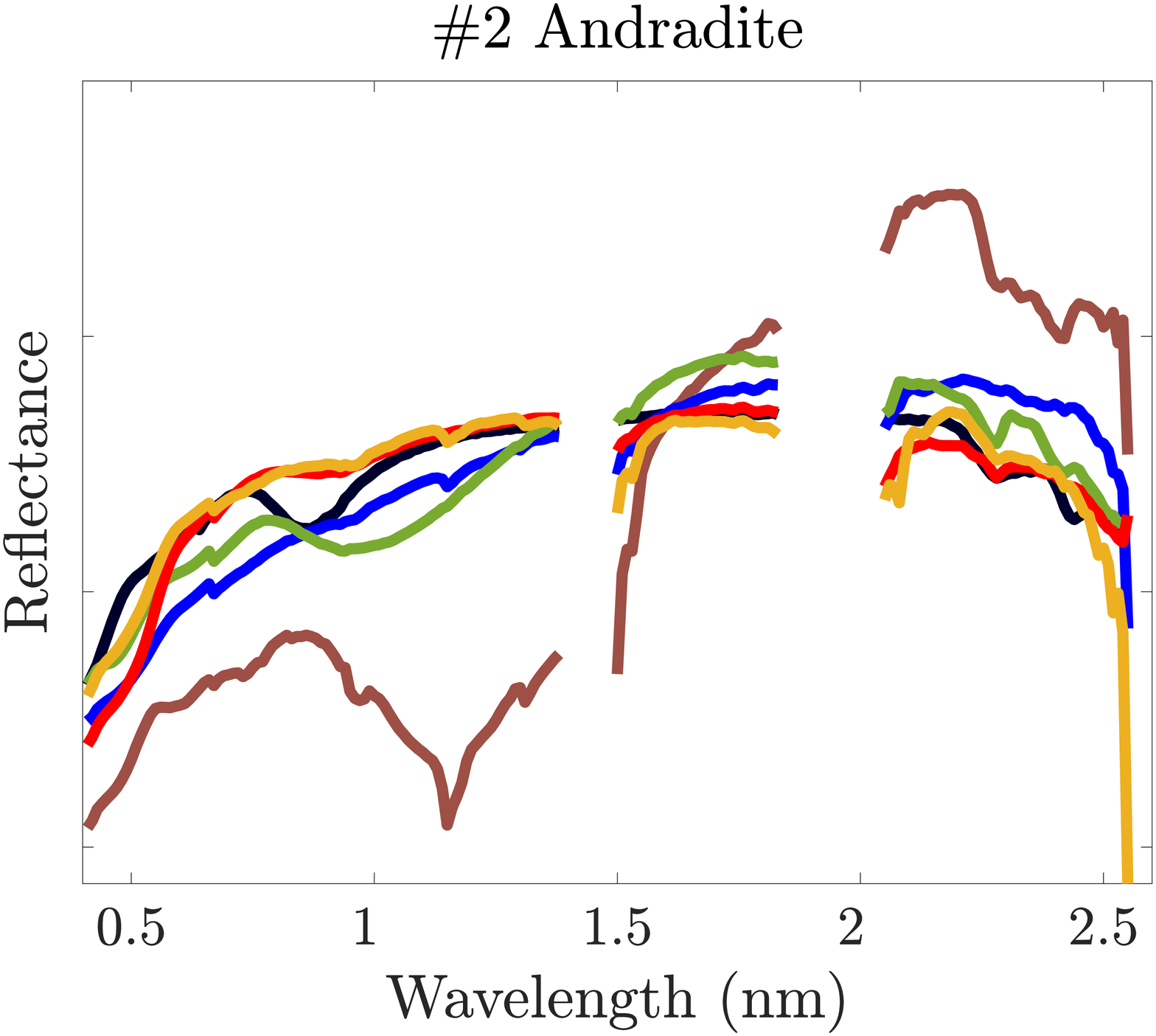}
		\includegraphics[width=0.30\linewidth]{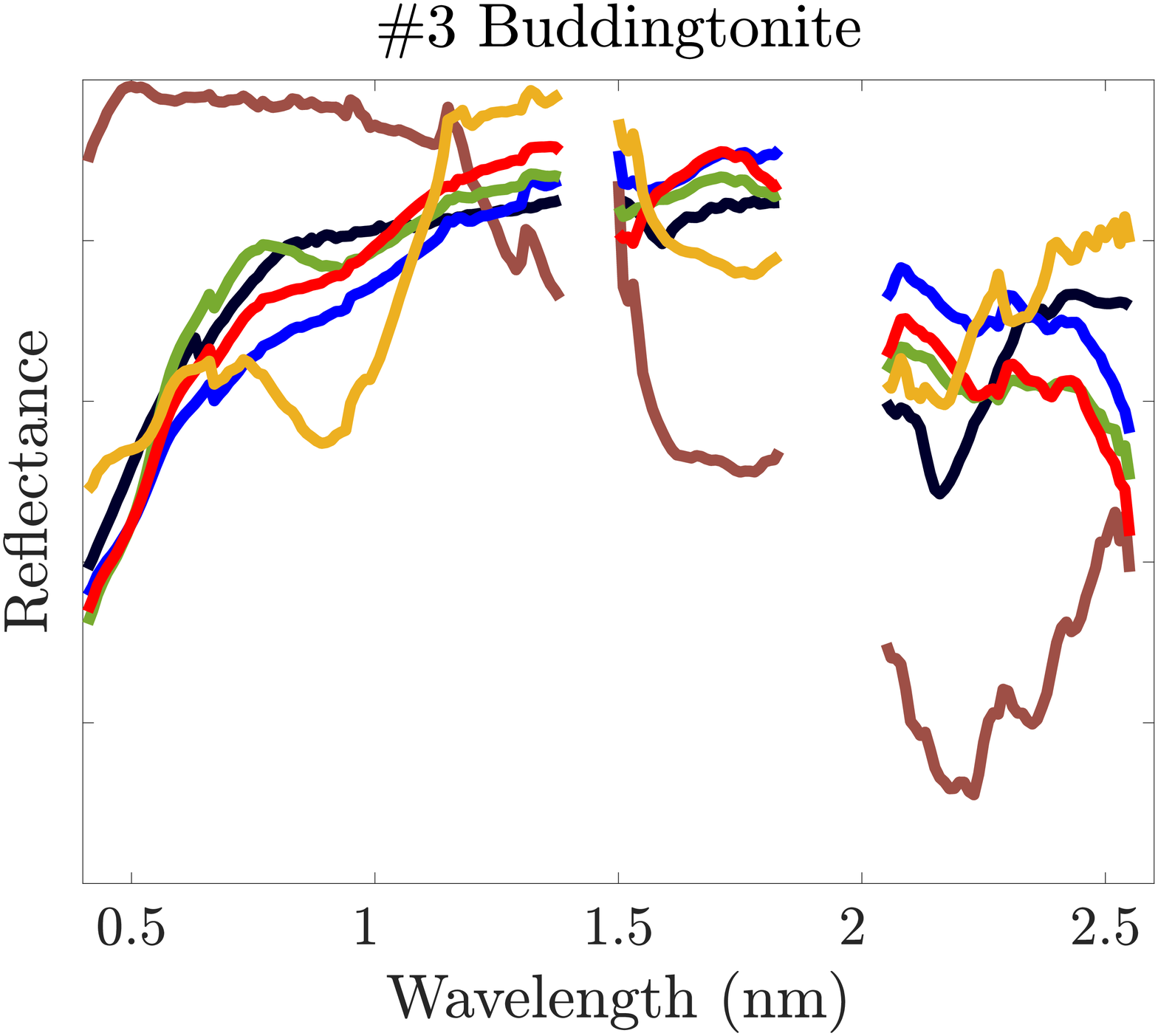}
		\vspace{0.1in}
	\end{minipage}
	\begin{minipage}[b]{\textwidth}
		\centering
		\includegraphics[width=0.30\linewidth]{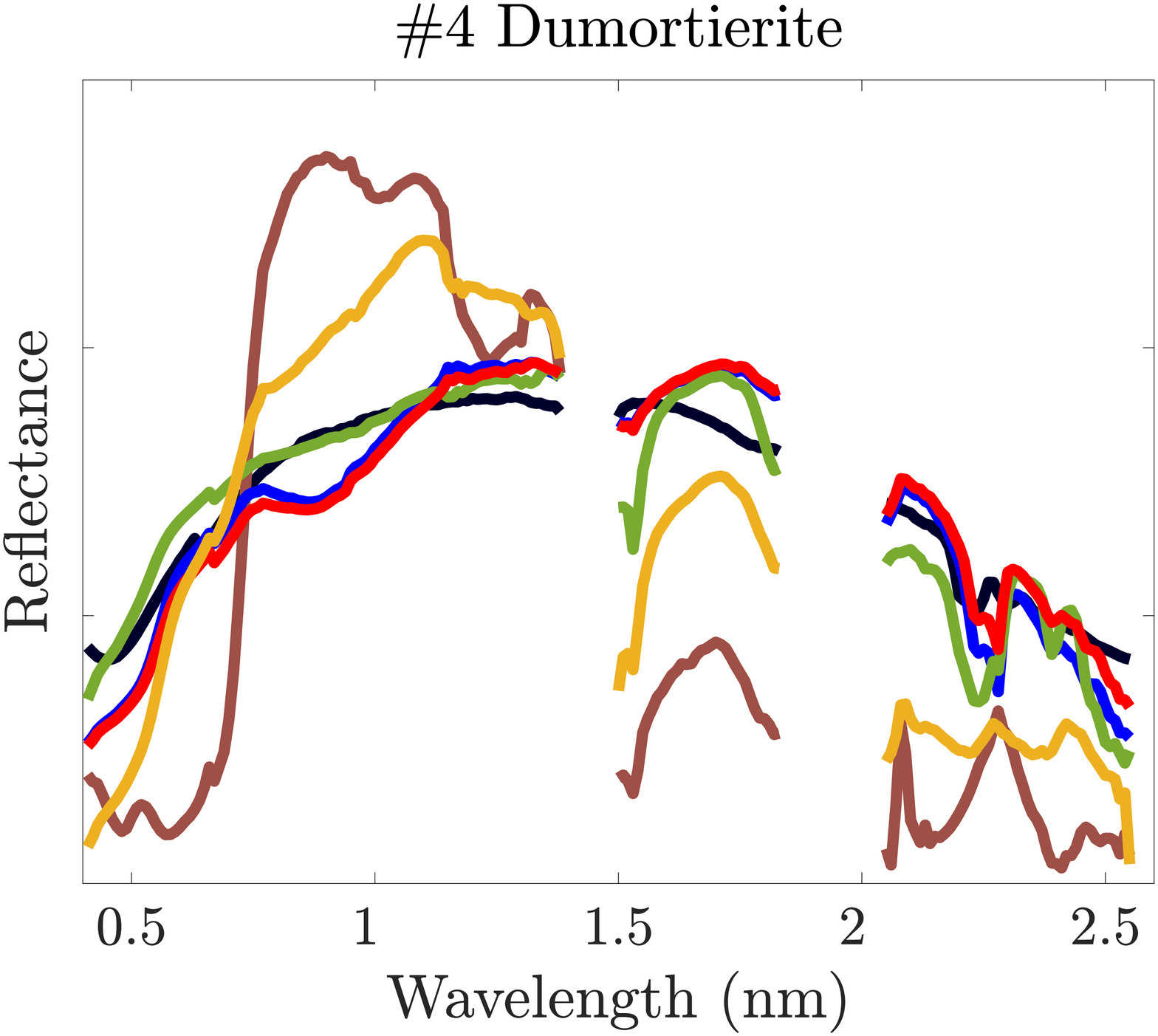}
		\includegraphics[width=0.30\linewidth]{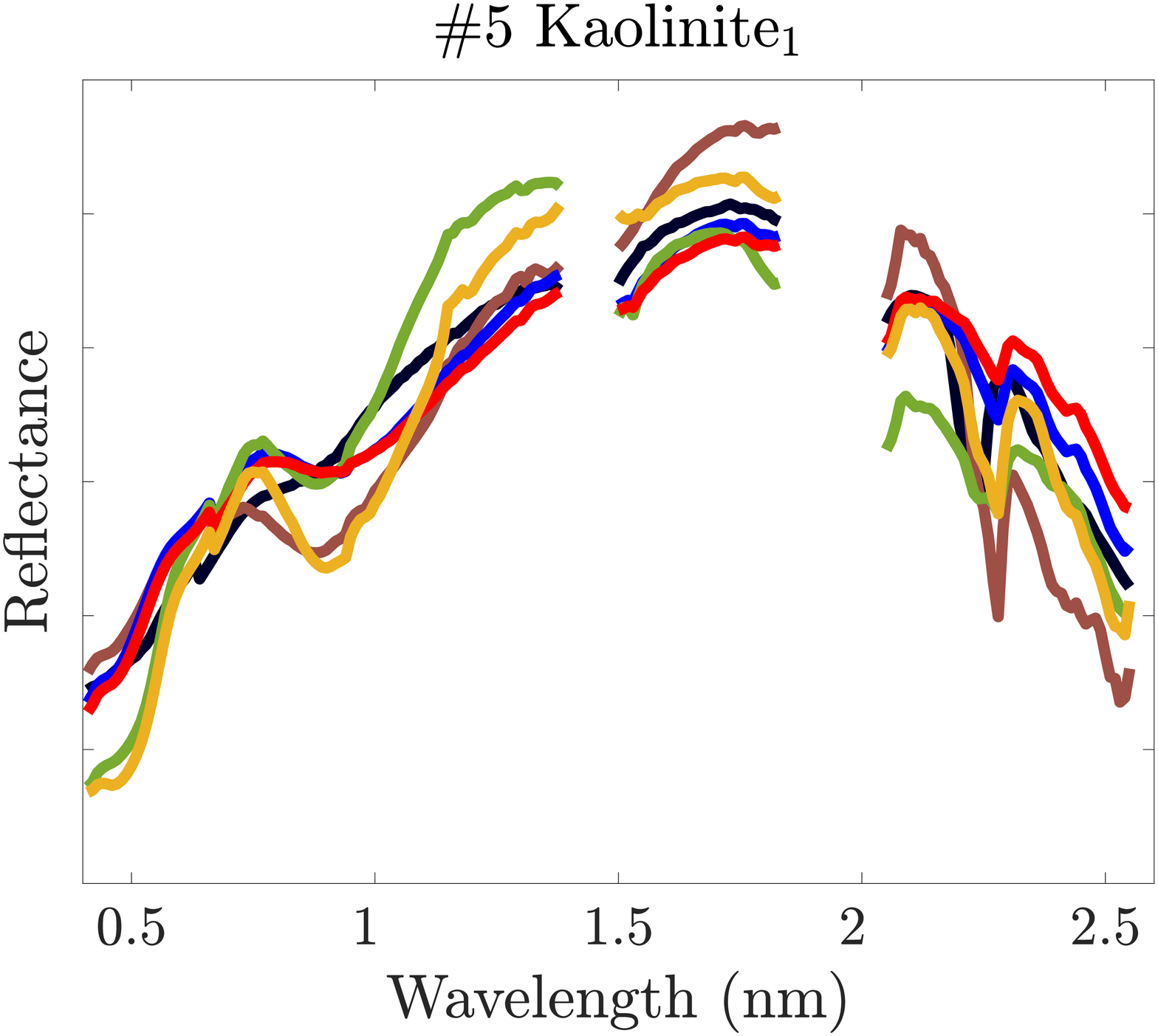}
		\includegraphics[width=0.30\linewidth]{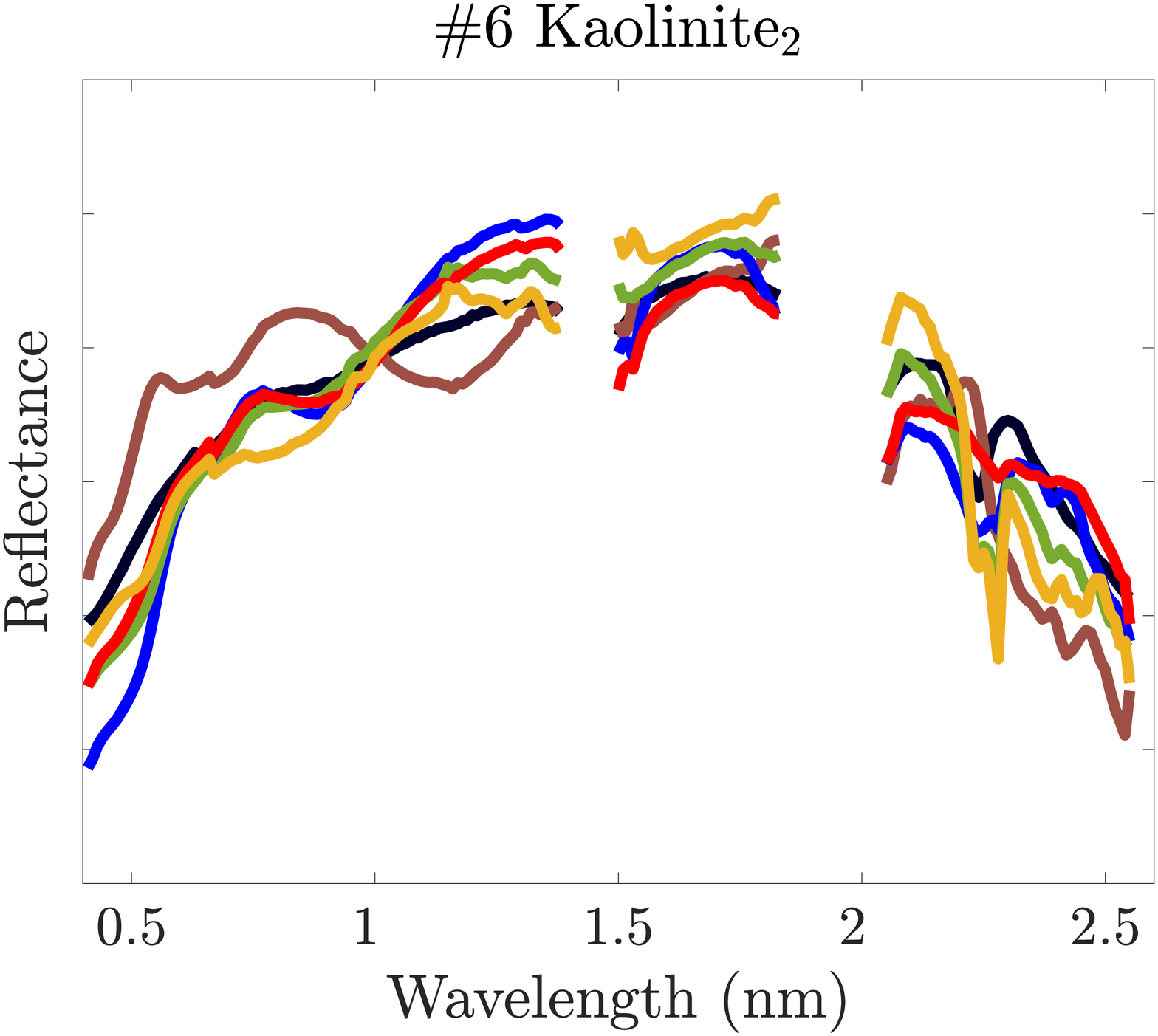}
		\vspace{0.1in}
	\end{minipage}
	\begin{minipage}[b]{\textwidth}
		\centering
		\includegraphics[width=0.30\linewidth]{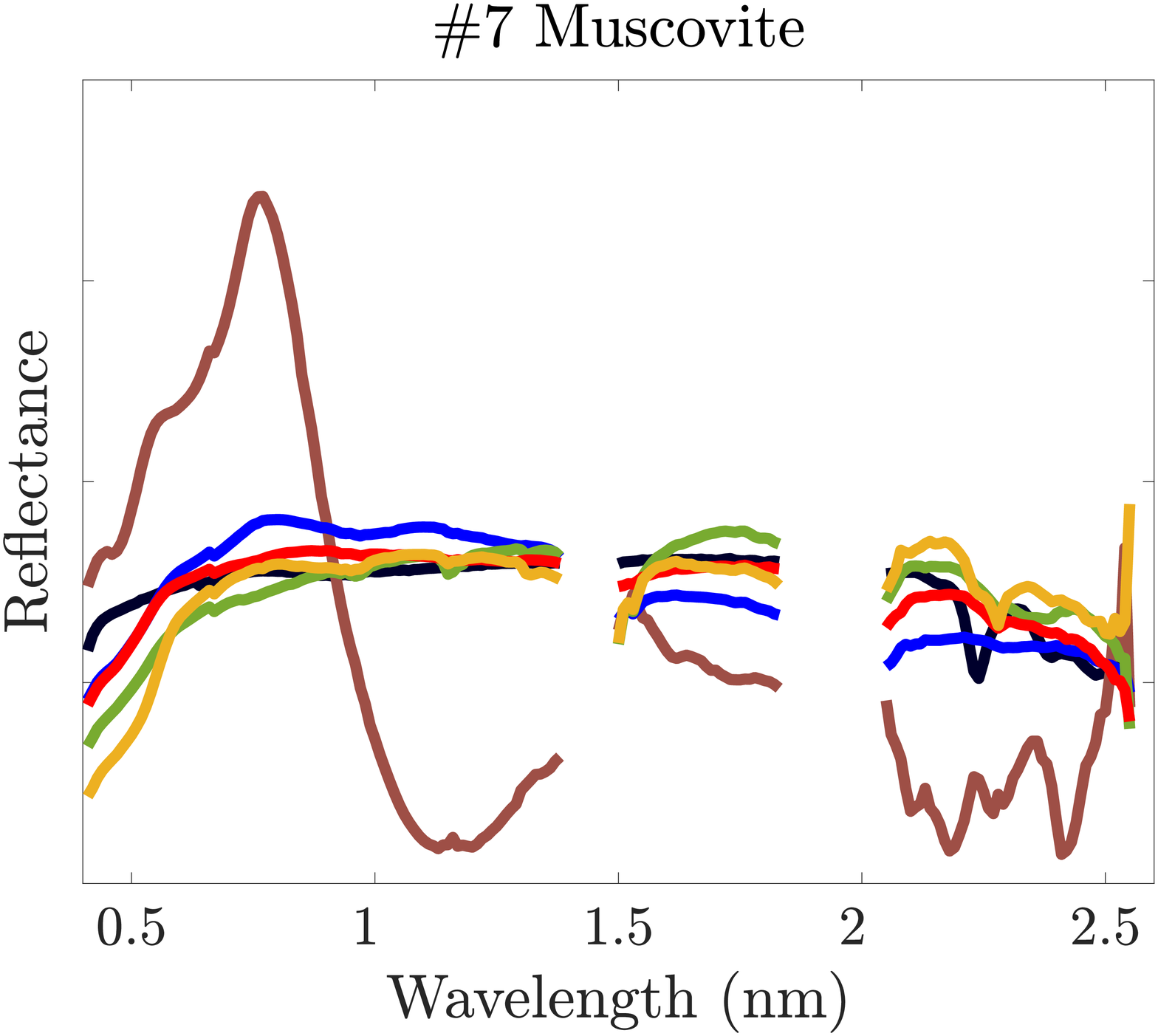}
		\includegraphics[width=0.30\linewidth]{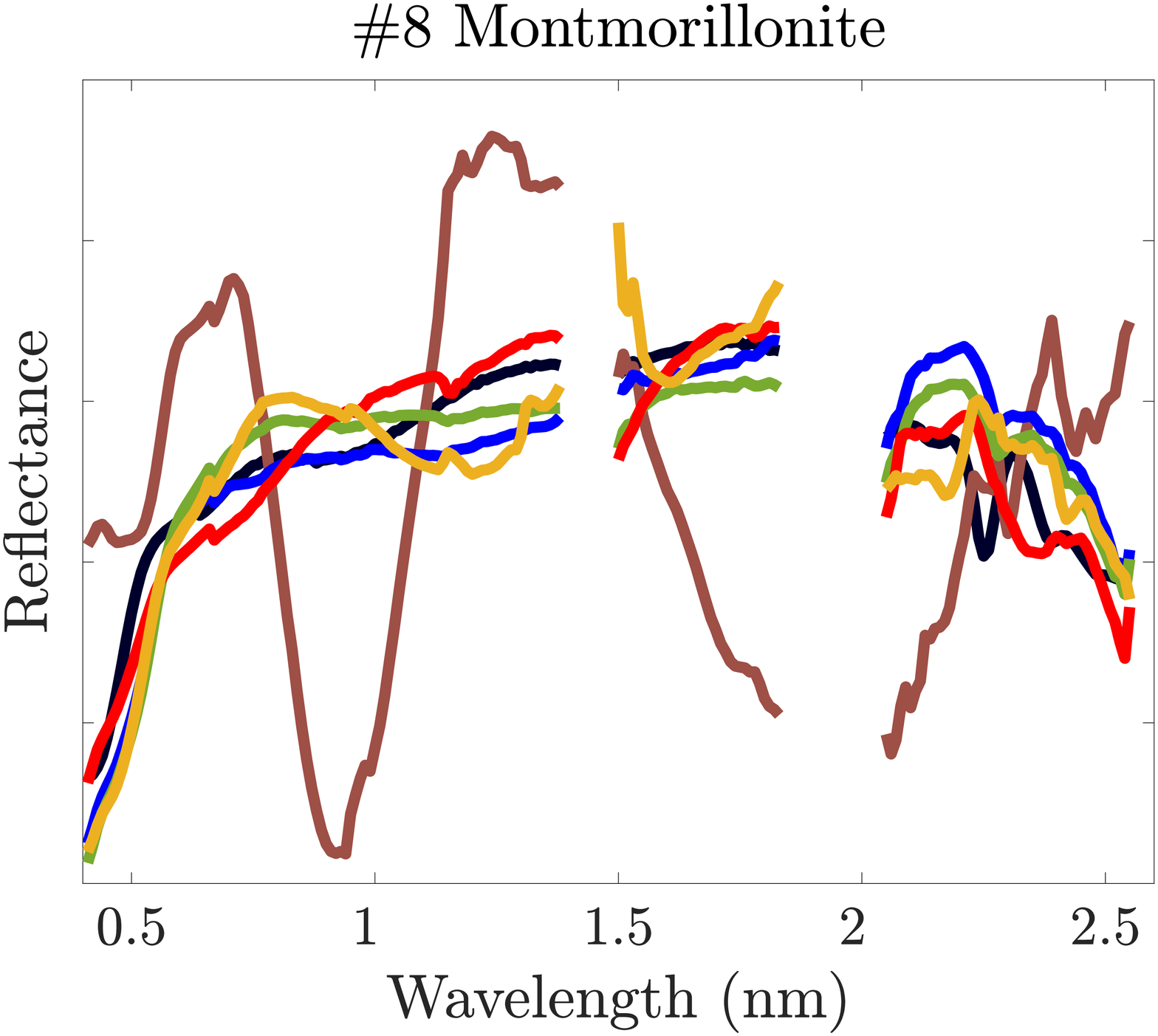}
		\includegraphics[width=0.30\linewidth]{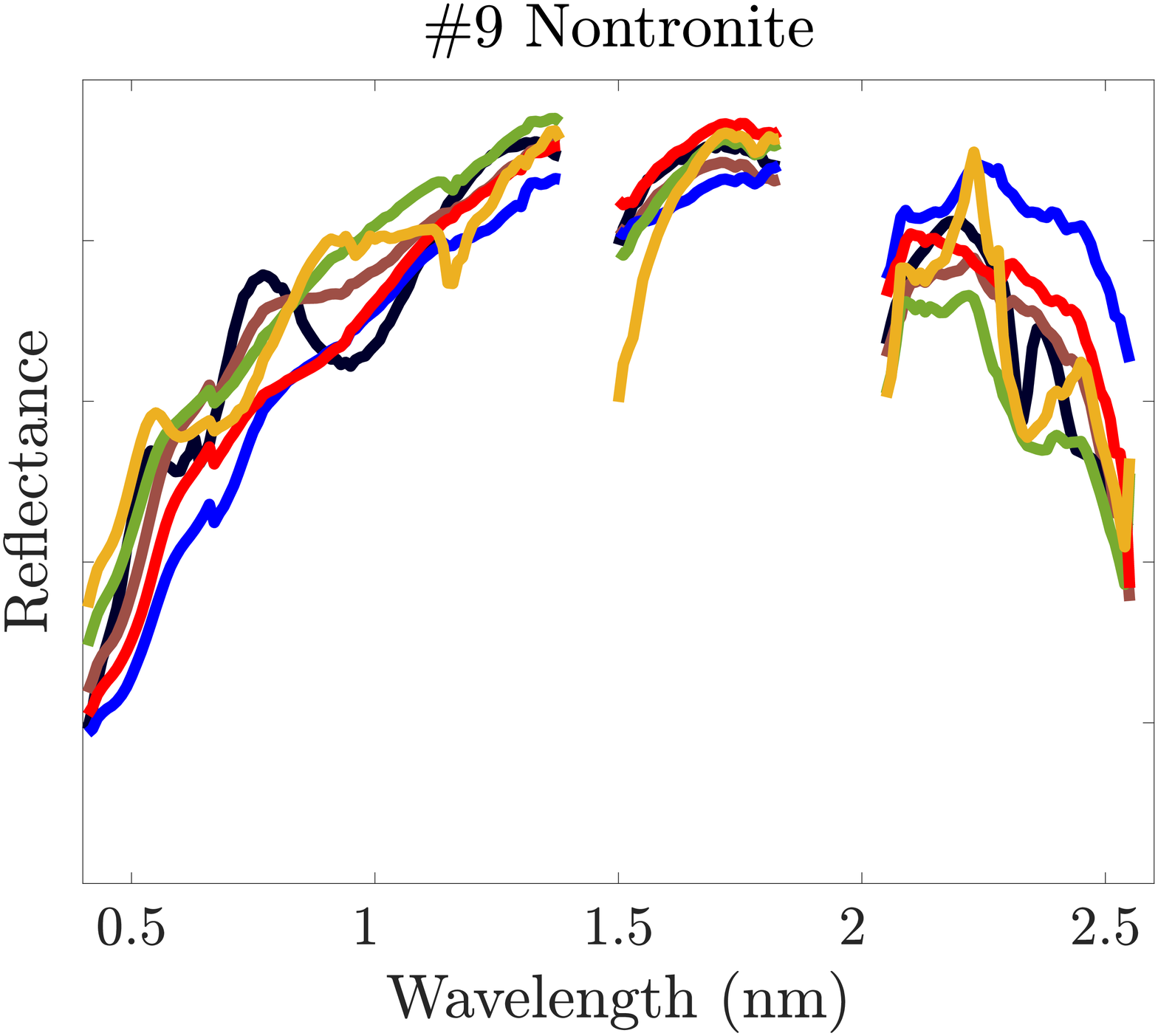}
		\vspace{0.1in}
	\end{minipage}
	\begin{minipage}[b]{\textwidth}
		\centering
		\includegraphics[width=0.30\linewidth]{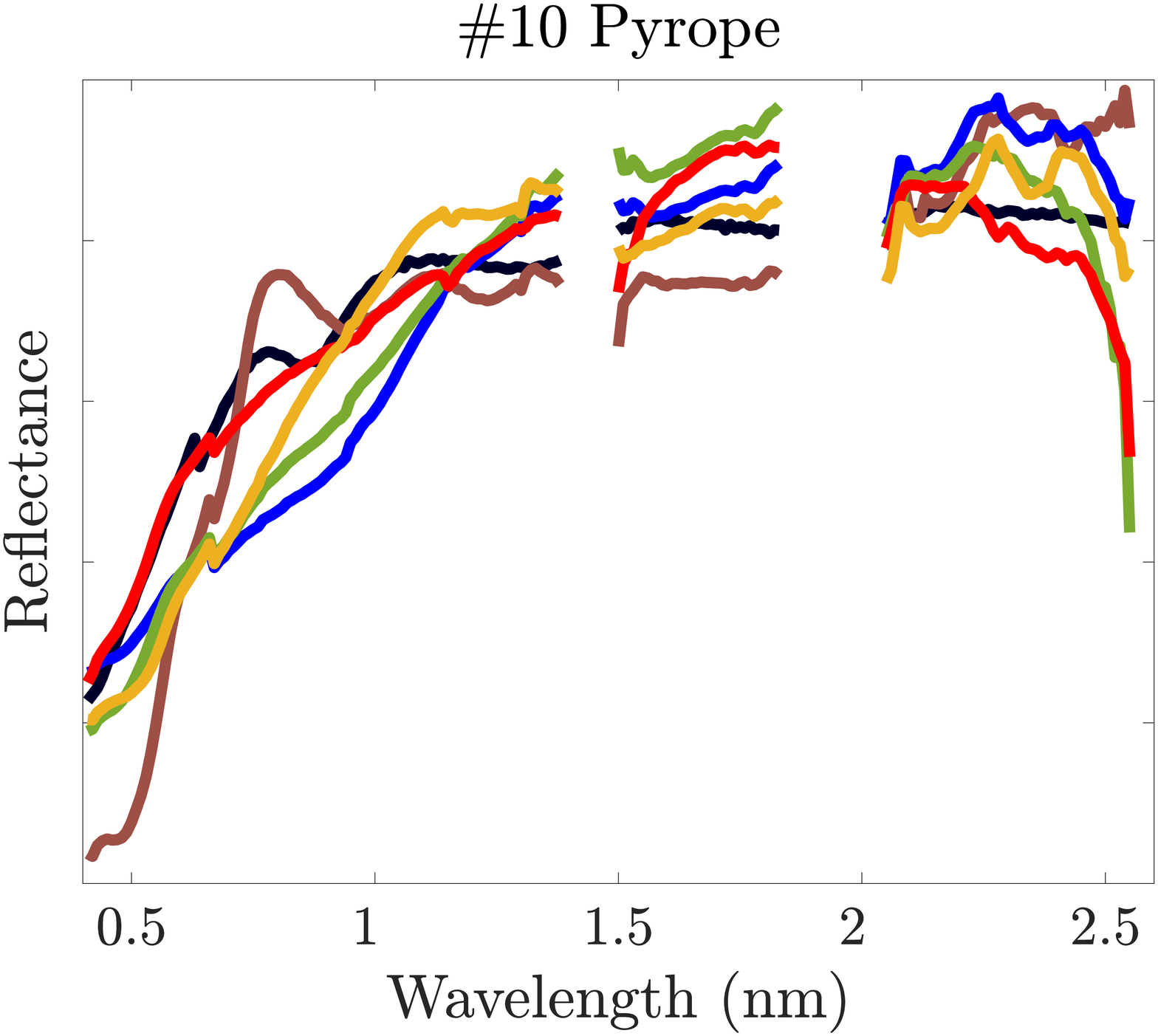}
		\includegraphics[width=0.30\linewidth]{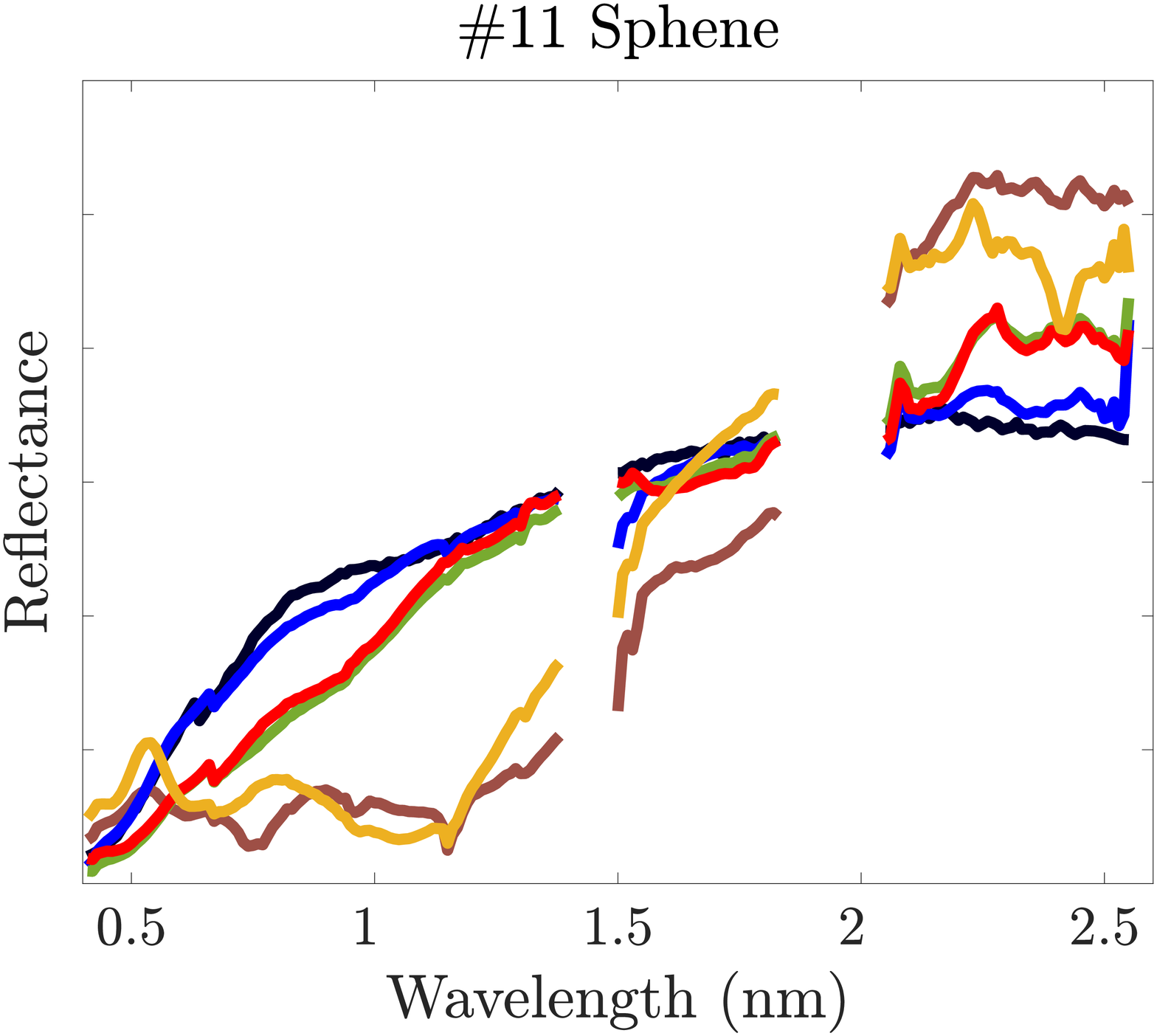}
		\includegraphics[width=0.30\linewidth]{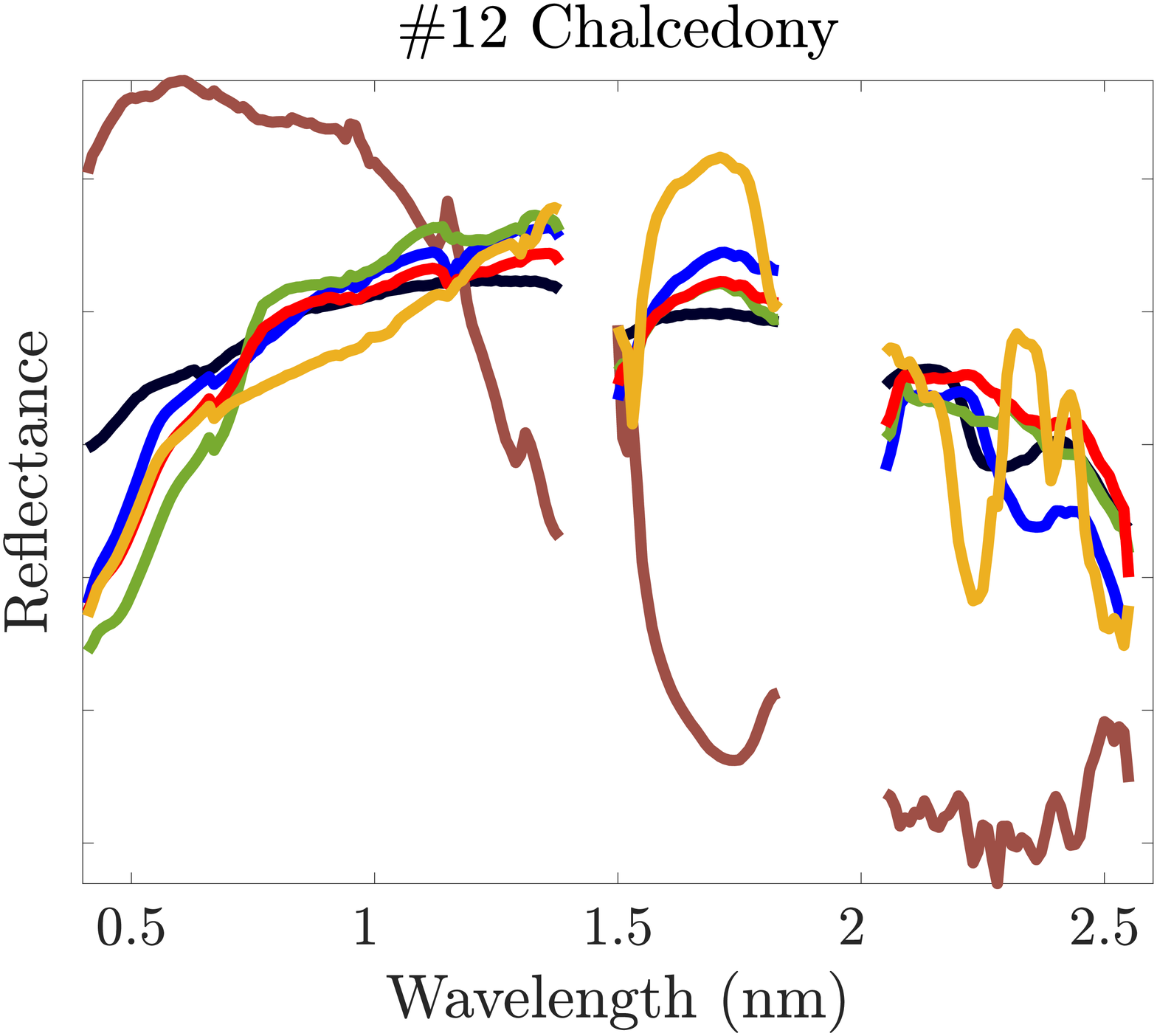}
	\end{minipage}
	\includegraphics[width=0.75\linewidth]{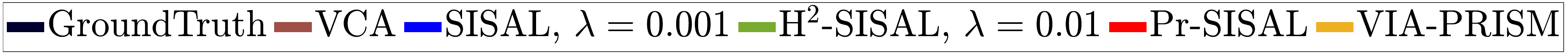}
	\caption{Estimated spectrums of Cuprite. Algorithms: VCA, SISAL with $ \lambda = 0.001 $, H$^2$-SISAL with $ \lambda = 0.01 $, Pr-SISAL, and VIA-PRISM.}\label{fig:Outlier}
\end{figure}

Table~\ref{Tab:CupriteExpOutliers} displays the SAD performance of the tested algorithms for $10$ trials (The locations and spectral responses of the outliers are changed at each trial).
It is seen that VCA gives the worst average SAD, which suggests that VCA is sensitive to outliers. The other algorithms, including the new possibility of H$^2$-SISAL and Pr-SISAL, are more robust as indicated by their SAD performance.
Fig.~\ref{fig:Outlier} shows the estimated spectral signatures $ \hat{\ba}_i $ of the various materials from one random trial. We observe that SISAL, H$^2$-SISAL and Pr-SISAL yield good recovery; VCA and VIA-PRISM are not as promising in comparison.

\begin{table}[!hbt]
	\centering
	\caption{SAD performances on the Cuprite dataset with outliers. The best SADs averaged over 10 trials among all the tested algorithms are marked in {\bf bold}.}
	\scalebox{0.75}{
		\begin{tabular}{c||c|c|c|c|c|c|c}
			\hline \hline
			\multirow{2}{*}{\diagbox{\small{Endmember}}{\small{Alg.}}}	&  \multirow{2}{*}{VCA} & \multicolumn{2}{c|}{SISAL} & \multicolumn{2}{c|}{H$^2$-SISAL} & \multirow{2}{*}{Pr-SISAL} & \multirow{2}{*}{VIA-PRISM}\\
			\cline{3-6}
			&   & ${\lambda = 0.001} $ & ${\lambda =0.01} $ & ${\lambda =0.01} $ & ${\lambda = 0.1} $ & &  \\
			\hline \hline
			Alunite		   	&  9.64$\pm$4.59 & 4.74$\pm$0.26  		   & 6.72$\pm$1.21  & \textbf{2.82}$\pm$1.30  & 5.84$\pm$1.50  &  3.91$\pm$0.77 		  & 11.65$\pm$2.72 \\
			\hline
			Andradite 	   	&  8.38$\pm$5.21 & 3.45$\pm$0.48  		   & 7.50$\pm$1.97  & 2.95$\pm$0.61  		  & 6.16$\pm$0.96  &  \textbf{2.27} $\pm$0.31 & 3.31$\pm$0.41 \\
			\hline
			Buddingtonite 	& 13.42$\pm$4.14 & 4.07$\pm$1.12  		   & 3.93$\pm$0.56  & \textbf{3.23}$\pm$0.69  & 5.49$\pm$0.90  &  3.47$\pm$0.31 		  & 3.85$\pm$1.02 \\
			\hline
			Dumortierite 	& 12.43$\pm$3.74 & \textbf{2.93}$\pm$0.83  & 6.51$\pm$1.31  & 3.17$\pm$0.52		      & 5.38$\pm$0.79  &  3.17$\pm$0.54 		  & 6.85$\pm$2.87 \\
			\hline
			Kaolinite$_1$  	&  9.00$\pm$4.05 & \textbf{2.33}$\pm$0.43  & 4.42$\pm$1.48  & 3.18$\pm$0.72  		  & 5.41$\pm$1.01  &  2.39$\pm$0.28 		  & 4.38$\pm$1.47 \\
			\hline
			Kaolinite$_2$ 	&  7.33$\pm$4.86 & 2.53$\pm$0.75  		   & 5.39$\pm$2.09  & 2.59$\pm$0.56  		  & 5.52$\pm$1.57  &  \textbf{2.34} $\pm$0.59 & 3.36$\pm$1.06  \\
			\hline
			Muscovite 	   	& 15.40$\pm$5.50 & \textbf{3.11} $\pm$0.59 & 5.14$\pm$2.25  & 3.66$\pm$1.24 		  & 5.30$\pm$1.32  &  3.25$\pm$0.58 		  & 4.57$\pm$0.64  \\
			\hline
			Montmorillonite & 10.31$\pm$3.65 & 3.47$\pm$0.57  		   & 3.29$\pm$0.24  & 2.31$\pm$0.91  		  & 3.42$\pm$0.53  &  \textbf{2.11} $\pm$0.48 & 2.79$\pm$0.28  \\
			\hline
			Nontronite 		& 5.92$\pm$2.96  & 3.66$\pm$0.57  		   & 3.75$\pm$0.64  & 3.33$\pm$0.98  		  & 4.46$\pm$1.03  &  \textbf{2.58} $\pm$0.42 & 3.36$\pm$0.73  \\
			\hline
			Pyrope 			& 12.59$\pm$3.87 & 2.72$\pm$1.00  		   & 5.79$\pm$2.17  & 3.44$\pm$0.89  		  & 5.15$\pm$1.45  &  \textbf{2.62} $\pm$0.53 & 3.11$\pm$0.65  \\
			\hline
			Sphene 			& 11.96$\pm$1.34 & \textbf{2.35}$\pm$0.91  & 5.91$\pm$1.37  & 2.99$\pm$0.62  		  & 6.30$\pm$2.09  &  3.69$\pm$0.66 		  & 9.85$\pm$1.39  \\
			\hline
			Chalcedony 		& 14.61$\pm$4.89 & 2.68$\pm$0.40  		   & 4.96$\pm$2.06  & 2.98$\pm$0.86  		  & 5.78$\pm$1.11  &  \textbf{2.58} $\pm$0.78 & 6.27$\pm$4.76  \\
			\hline\hline
			Average SAD   	& 10.91          & 3.17  		           & 5.28 			& 3.05 					  & 5.35 		   & \textbf{2.86} 			  & 5.28  \\
			\hline
	\end{tabular}}
	\label{Tab:CupriteExpOutliers}
\end{table}

\section{Conclusions}
\label{sec:conclusions}

In this article we showed that the famous SISAL algorithm, developed by Bioucas-Dias in hyperspectral unmixing in 2009, can be explained as a probabilistic method for SCA.
In particular, SISAL was derived from the noiseless case, and our study provides an explanation of why SISAL can be robust to noise.
Moreover, we gave a positive answer to the question of whether the SISAL algorithm can lead to provable convergence to a stationary point.
This was done by casting SISAL as an instance of a proximal gradient framework in non-convex first-order optimization.
Furthermore, through connecting SISAL and probabilistic SCA, we also found new SCA formulations that resemble SISAL.
To allow us to numerically study the new SCA formulations, we built customized algorithms for them.
The potential of the new algorithms was demonstrated by numerical experiments.

\bibliographystyle{IEEEtran}
\bibliography{refs}



\clearpage
\appendix
\section*{Appendix}
\subsection*{A.~~Additional Simulation Results}
\label{sect:add_sim}

\begin{figure}[!hbt]
	\centering
	\includegraphics[width=.75\linewidth]{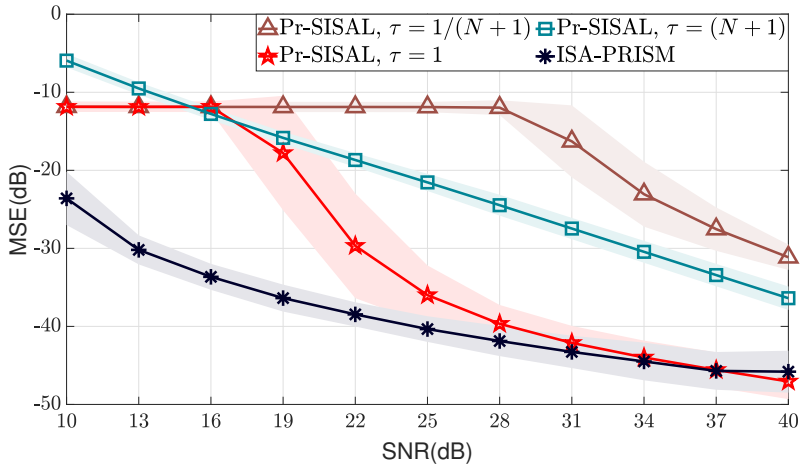}
	\caption{Performance of the formulation in \eqref{eq:Form2_lambda} for different values of $\tau$.}
	\label{fig:Pr_SISAL_Variants}
\end{figure}

We display two more numerical results for Pr-SISAL.
The first is with Heuristic 1, which is used to build the approximate ML formulation in Formulation 3.
To put into context, let us write down a slightly more general form of Formulation 3:
\beq  \label{eq:Form2_lambda}
\min_{\bB^\top \bone = \bp}   ~ - \log ( | \det(\bB) | ) -  \frac{\tau}{T} \sum_{t=1}^T \sum_{i=1}^N \log \Phi \left( \frac{ \bb_i^\top \by_t}{ \sigma \| \bb_i \| } \right),
\eeq
where $\tau > 0$, and Formulation 3 is the special case of $\tau = 1$.
In Remark 2, we argue that $\tau= 1/(N+1)$ is arguably equipped with a better rationale (lower-bound approximation of the ML objective), but eventually the heuristic (and, intuitively, more progressive) choice of $\tau =1$ prevails in terms of approximating the ML problem better in practice.
We want to illustrate that.
Fig.~\ref{fig:Pr_SISAL_Variants} shows the performance of formulation in \eqref{eq:Form2_lambda} for different values of $\tau$ and for $(M,N)= (10,5)$, $T=1,000$;
the simulation is done by exactly the same way as in Section~\ref{sect:syth_sim}.
We see that $\tau = 1/(N+1)$ does not work well, except for very high SNRs.
We also try $\tau =N+1$ (more progressive than $\tau = 1$), and the result is not as good as $\tau= 1$.

\begin{figure}[!hbt]
	\begin{minipage}[a]{.6\linewidth}
		\centering
		\includegraphics[width=.96\linewidth]{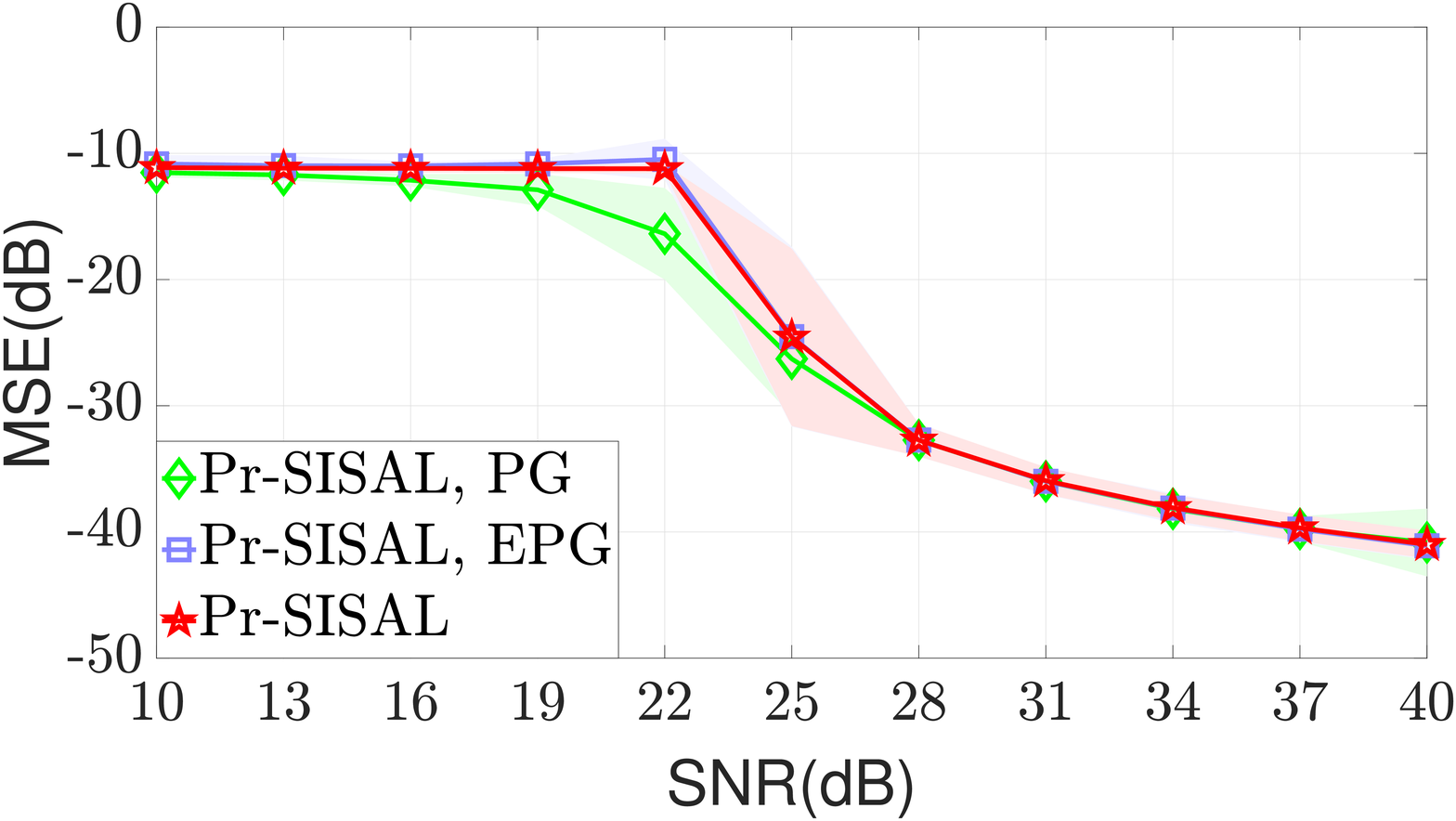}
		\figcaption{Performance comparison of several Pr-SISAL implementations.}
		\label{fig:SV_vs_PG}
	\end{minipage}%
	\begin{minipage}[b]{.4\linewidth}
		\centering
		\tabcaption{Average runtime (in sec.) for several Pr-SISAL implementations.}
		\small
		\renewcommand{\arraystretch}{1.2}
		\begin{tabular}{c||c}
			\hline \hline
			Algorithms  & Runtimes \\
			\hline \hline
			Pr-SISAL, PG  & 198.814  \\
			\hline
			Pr-SISAL, EPG & 243.307  \\
			\hline
			Pr-SISAL      & 21.542 \\
			\hline\hline
		\end{tabular}
		\label{Tab:SV_vs_PG}
	\end{minipage}
\end{figure}

The second result is about the implementations of Formulation 3.
It was mentioned that the proximal gradient method can be used to handle Formulation 3, but the results are not promising.
Here we show the results.
We implement Formulation 3 using the same proximal gradient algorithm in Algorithm~\ref{alg:h2sisal}, with or without extrapolation.
We stop the algorithm if ${\rm rc}(\bB^{k+1},\bB^k) \leq 10^{-8}$ or if the number of iterations exceeds $4 \times 10^5$.
Fig.~\ref{fig:SV_vs_PG} and Table~\ref{Tab:SV_vs_PG} show the MSE and runtime performance, respectively,
for $(M,N,T)= (20,10)$, $T=1,000$;
the simulation settings are the same as the previous.
There, ``Pr-SISAL'', ``Pr-SISAL, PG'' and ``Pr-SISAL, EPG'' refer to the inexact BCD algorithm in Algorithm~\ref{alg:prsisal}, the proximal gradient algorithm and the extrapolated proximal gradient algorithm, all for Formulation 3.
We see that all the implementations yield similar MSE performance, but the proximal gradient implementations are very slow.

\end{document}